\documentclass[jgrga]{AGUTeX2}
\usepackage{graphics}
\usepackage{multirow}
\usepackage{graphicx}
\usepackage{enumerate}
\usepackage{amsmath}
\usepackage{amsfonts}
\usepackage{epsfig}
\usepackage{color}
\usepackage{rotating}
\usepackage{float}

\newcommand{\bb}{{\textbf b}}

\newcommand{\br}{B_r}
\newcommand{\bu}{{\textbf u}}

\newcommand{\MM}{\mathcal{M}}

\newcommand*{\rom}[1]{\expandafter\@slowromancap\romannumeral #1@}

\DeclareMathOperator*{\argmax}{arg\,max}

\authorrunninghead{BAERENZUNG, HOLSCHNEIDER, WICHT, SANCHEZ, LESUR}

\titlerunninghead{}

\begin{document}
\title{Modeling and predicting the short term evolution of the Geomagnetic field}
\authors{Baerenzung Julien,\altaffilmark{1}
Holschneider Matthias,\altaffilmark{1} Wicht Johannes, \altaffilmark{2}  Sanchez Sabrina, \altaffilmark{2} Lesur Vincent,\altaffilmark{3}}
\altaffiltext{1}{Institute for Mathematics, University of Potsdam, Potsdam, Germany}
\altaffiltext{2}{Max Planck Institute for solar system research, G\"ottingen, Germany}
\altaffiltext{3}{Institut de Physique du Globe de Paris, Paris, France}
\begin{abstract} 
\noindent The coupled evolution of the magnetic field and the flow at the Earth's core mantle boundary is modeled within the $1900.0-2014.0$ time period. 
To constraint the dynamical behavior of the system with a core field model deriving from direct measurements of the Earth's magnetic field 
we used an Ensemble Kalman filter algorithm.
By simulating an ensemble of possible states, access to the complete statistical properties of the considered fields is available. Furthermore,
the method enables to provide predictions and to assess their reliability. 

\noindent In this study, we could highlight the cohabitation of two distinct flow regimes.
One associated with the large scale part of the eccentric gyre, which evolves slowly in time and posses a very long memory of its past, and 
a faster one associated with the small scale velocity field. We show that the latter can exhibit rapid variations 
in localized areas. The combination of the two regimes  can  
predict quite well the decadal variations in length of day, but it can also explain the discrepancies between the physically predicted and the
observed trend in these variations.

\noindent Hindcast tests demonstrate that the model is well balanced and that it can provide accurate short term predictions of 
a mean state and its associated uncertainties. However, magnetic field predictions are limited by the high 
randomization rate of the different velocity field scales, and after approximately $2000$ years of forecast, no reliable information
on the core field can be recovered.
\end{abstract}
\begin{article}

\section{Introduction}

\noindent  In the Earth's outer core, turbulent motions of the electrically conducting fluid
sustains the geomagnetic field through dynamo action. Part of this field, the poloidal one,
crosses the mantle and can be observed at the Earth's surface and above. Because of the low
conductivity of the mantle (see \citet{Velimsky2010,Jault2015}), once measured and modeled,
the poloidal field can be estimated everywhere outside and at the outer boundary of the core. 
Since at the core mantle boundary (CMB) the flow is advecting the magnetic field, a close examination
of the  geomagnetic field variations at the Earth's surface, make it possible to infer the velocity field
responsible for them.

\noindent Within the outer core, the evolution of the magnetic field is prescribed by the induction equation.
At the CMB this equation can be simplified. Under the assumption that the mantle is a perfect
electrical insulator, the toroidal field, which interacts with the poloidal field inside
the outer core, vanishes at the CMB. In addition, since the fluid cannot penetrate the mantle, its associated velocity field
is purely two dimensional. And finally, on short timescales, diffusion effects can be considered as negligible in comparison 
to advection effects as shown in \citet{Holme2007}. So all in all, the induction equation expressed at the CMB can be simplified 
into the so-called  Frozen Flux (FF) approximation  (see \citet{Backus1996}). By inverting this equation,
which couples the velocity field to the radial component of the magnetic field and
the secular variation, fluid motions at the CMB can be recovered. 

\noindent However, since the velocity field has two components for one equation, and since any flow scale can interact 
with the magnetic field to generate the large scale, observable, secular variation (SV), the inverse problem is ill-posed.
To reduce the non-uniqueness of the velocity field, physical assumptions decreasing the dimension
of possible solutions can be made. Most of the constraints generally used in core flow inversions, such as 
quasi-geostrophy, columnar, tangential-geostrophy or purely toroidal, are derived in \citet{Holme2007,Finlay2010}.  
Nevertheless, to obtain a unique solution, additional constraints on the velocity field need to be imposed. 
Typically, one enforces the energy associated with the small scale velocity field to rapidly decay,
based on the so-called large scale assumption (see \citet{Holme2007,Finlay2010}). However,
\citet{Baerenzung2016} has shown that although the flow is dominant at large scales, its total kinetic energy spectrum does not exhibit 
a strong decaying slope. 
Recently, other strategies have been developed to bypass the issues raised by the nonuniqueness of the velocity field. 
In particular, \citet{Aubert2014} proposed to use the statistical properties of 
an Earth like geodynamo simulation (the Coupled-Earth model of \citet{Aubert2013}) to characterize the flow and the magnetic field a priori.
A major advantage of such an approach is that the correlations between the fields at the CMB and the fields within the outer
core are available allowing for imaging the entire outer core state.

\noindent Constraining a priori the temporal dependency of the velocity field is a more delicate operation as 
constraining it spatially. Optimally, one should account for the dynamics of the outer core fluid and magnetic field prescribed by the 
magnetohydrodynamic equations.  Approaches such as variational data assimilation (see \citet{Canet2009,Li2014}) allow for physical
modeling within an inversion framework. To do so, the method search for
the optimal initial conditions of the system in order for the deterministic trajectories of the different fields to explain at best
the observations. The drawback of the method is that all data are treated simultaneously, so whenever the dimension of data or 
their amount are large,
the algorithm become computationally expensive. An alternative to avoid such a block inversion is to operate sequentially.
\citet{Kuang2008} were the first to adopt a sequential assimilation algorithm in the context of geomagnetic modeling.  
The optimal interpolation algorithm they used proceeds recursively in two steps. In the first one, the state variables are propagated
in space and time with a given physical model,  in the case of \citet{Kuang2008}, a three-dimensional geodynamo simulation.
Once observations become available, the state variables are corrected accordingly, and then the prediction restart.
Since with this approach uncertainties are not modeled, they have to be specified in an ad-hoc manner.
Algorithms which permit to simulate a mean model and its associated uncertainties exist, the Kalman filter (KF) approach 
(see \citet{Kalman1960,Talagrand1997,Cohn1997,Evensen2003})
is one them. As for the optimal interpolation algorithm the KF proceeds sequentially with
a forecast and an analysis steps. The main difference is that the evolution of errors
is also predicted, and  whenever data become available, these errors are taken into account for the Bayesian update of the state variables. 
The KF can only be applied to systems exhibiting a linear dynamics. When the dynamics of the system
is nonlinear, as it is the case here for the geomagnetic field, propagation of errors cannot be analytically derived. However,
it can be either approximated by linearization with the extended version of the KF, or represented through an ensemble of possible solutions in the
ensemble Kalman filter (EnKF). As for \citet{Barrois2017} the latter option is the one we chose in this study.

\noindent In the EnKF, the different fields of interest are represented through an ensemble of possible states. For the
prediction step, the dynamical model of the system prescribes the spatio temporal evolution of each individual member of the ensemble. At
the analysis, covariances deriving from the forecasted fields and data are combined to correct the state of the ensemble predicted at observation
time. Due to the limitations in available computational power, a balance between complexity of the dynamical model and size of the ensemble
has to be found. Here we decided to favor accuracy in statistical representation. Therefore, the evolution of the core  magnetic field 
and velocity field are only modeled at the level of the core-mantle boundary, through respectively the frozen flux equation
and a first order auto regressive process. Extending the approach of \citet{Baerenzung2016} to the time domain,
the parameters of the auto regressive process, are assumed to derive from scale dependent power laws, and are directly estimated
with the COV-OBS.x1 core magnetic field secular variation model of \citet{Gillet2013,Gillet2015b}.

\noindent The article is organized as follows. In section \ref{governingEquations} the mathematical approach chose to tackle the inverse problem
is described. It is then applied to the real geophysical context, and the results are shown in section \ref{results}. 
Finally some conclusions are drawn in section \ref{conclusion}.

\section{Models and parameters}\label{governingEquations}

\subsection{Quantities of interest and notations}\label{notations}

\noindent In this study three fields are of particular interest, the radial component of the magnetic field
$B_r(x,t)$, the secular variation $\partial_t B_r(x,t)$, and the velocity field $u(x,t)$ all expressed at the Earth's core mantle boundary.

\noindent The spectral counter parts of $B_r(x,t)$ and $\partial_t B_r(x,t)$ are respectively given by the spherical harmonics
coefficients $b_{l,m}$ and ${\gamma}_{l,m}$ such as:
\begin{eqnarray}
B_r(x,t) & = & -\sum_{l=1}^{l=+\infty} (l+1)\sum_{m=-l}^{m=+l}
b_{l,m}(t)Y_{l,m}(x) \ , \\
\partial_t B_r(x,t) & = & -\sum_{l=1}^{l=+\infty} (l+1)\sum_{m=-l}^{m=+l}
{\gamma}_{l,m}\left(t\right)Y_{l,m}(x) \  , 
\label{BrSHdecompostion}
\end{eqnarray}
with $Y_{l,m}(x)$ the Schmidt semi-normalized spherical harmonics (SH) of degree $l$ and order $m$.

\noindent The velocity field at the CMB $u(x,t)$ is decomposed into a poloidal  $\phi(x,t)$ and toroidal $\psi(x,t)$ scalar field
such as:
\begin{equation}
 u(x,t) = {r}\times {\nabla_H} \psi(x,t) + {\nabla_H} (|r|\phi(x,t)) \label{poloToro} \ ,
\end{equation}
where ${\nabla_H}$ corresponds to the horizontal divergence operator.
In spectral space, the poloidal and toroidal fields are respectively  deriving from  the coefficients $\phi_{l,m}$ and $\psi_{l,m}$ through the formulation:
\begin{eqnarray}
 \phi(x,t) & = & \sum_{l=1}^{l=+\infty} \sum_{m=-l}^{m=+l} \phi_{l,m}(t){Y}_{l,m}(x) \ ,\\ 
 \psi(x,t) & = & \sum_{l=1}^{l=+\infty} \sum_{m=-l}^{m=+l} \psi_{l,m}(t){Y}_{l,m}(x) \ .
\end{eqnarray}

\noindent According to the spherical harmonics expansion of $B_r(x,t)$, $\partial_t B_r(x,t)$ and $u(x,t)$,
we define the magnetic field and secular variation energy spectra as following:
\begin{eqnarray}
E_b(l) &=& (l+1)\sum_{m=-l}^{m=l} b_{l,m}^2  \label{magneticSpectrum}  \\
E_\gamma(l) &=& (l+1)\sum_{m=-l}^{m=l}\gamma_{l,m}^2 \label{svSpectrum} \ .
\end{eqnarray}
and the velocity field poloidal and toroidal energy spectra respectively as:
\begin{eqnarray}
E_\phi(l) &=& \frac{l(l+1)}{2l+1}\sum_{m=-l}^{m=l}\phi_{l,m}^2  \label{poloidalSpectrum}  \\
E_\psi(l) &=& \frac{l(l+1)}{2l+1}\sum_{m=-l}^{m=l}\psi_{l,m}^2 \label{toroidalSpectrum} \ .
\end{eqnarray}

\noindent In the following, we will also adopt two types of notation, one with normal characters and
another one with bold characters.
Normal characters will be associated with single epoch quantities. This includes $\gamma$, $b$ and $u$,
which will respectively contain the spherical harmonics coefficients of the secular variation
$\gamma_{l,m}$, the magnetic field $b_{l,m}$, and both the poloidal and toroidal
fields $\phi_{l,m}$ and $\psi_{l,m}$, but also any scalar, vector or matrix associated with a quantity
expressed at a given time. 

\noindent Bold characters will be used for quantities depending on both space and time.
As an example, a vector $\bf{a}$ will contain spatial vectors $a$ at $N$ different epochs, such as:
\begin{equation}
{\bf a}  = \left(a_0,a_1, \cdots, a_{N-2},a_{N-1}\right)^T
\end{equation}

\noindent Finally, a constant formalism for some statistical quantities will be used all over the manuscript.
The mean and maximum value of a distribution $p(a)$ will respectively be written with an over bar and a hat such as:
\begin{eqnarray}
 \bar{a} & = & E[a] = \int a p(a) da \\
 \hat{a} & = & \argmax_a \left(p\left(a\right)\right)
\end{eqnarray}
and the covariance associated with a random variable $a$ or between a random variable $a$ and a random variable $b$ will respectively be expressed as:
\begin{eqnarray}
 \Sigma_a & = & E\left[\left( a - \bar{a}\right) \left( a - \bar{a}\right)^T\right]   \\
 \Sigma_{ab} & = &  E\left[\left( a - \bar{a}\right) \left( b - \bar{b}\right)^T\right] \ .
\end{eqnarray}
Note that the use of bold characters for space time dependent variables also applies to the latter statistical quantities.

\subsection{Characterization of the magnetic field dynamical behavior}

\noindent As mentioned in the introduction, under the assumption that the mantle is a perfect electrical insulator, and supposing that
the observed secular variation is mainly induced by advection of the magnetic field at the CMB, the dynamical evolution
for the radial component of the magnetic field $Br$ at the CMB is given by the frozen flux equation which reads:
\begin{equation}
 \partial_t \br\left(x,t\right) = -{\nabla_H}\left(u\left(x,t\right) \br\left(x,t\right)\right) \ .\label{FF}
\end{equation}
\noindent Following the notations given in section \ref{notations}, this equation  can be written in spectral space as:
\begin{equation}
 \gamma =  -A_b u = -A_u b = -A(ub)\ .\label{FFv}
\end{equation}
where the linear operators $A_b$ and $A_u$ and the third order tensor $A$ allow us to calculate the SH coefficients
associated with the advection term ${\nabla_H}(u\left(x,t\right) \br\left(x,t\right))$ when they are respectively applied to $u$, $b$
and $(ub)$.

\subsection{Characterization of the flow spatio-temporal behavior}
\noindent The dynamical evolution of the fluid within the Earth's outer core is prescribed by the Navier-Stokes equations. 
Numerically solving this equation is not only numerically expensive, it remains out of reach at the regime the Earth is exhibiting.
Since on short timescales, the observable magnetic field and secular variation only depend on the velocity field at the CMB, we propose
to model the outer core flow only at this location. Following \citet{Gillet2015} we chose a first order autoregressive process
to do so.
With such a model, the velocity field $u(t+\Delta t)$, depends on the velocity field $u(t)$ through the relation:
\begin{equation}
 u(t+\Delta t) = \Gamma(\Delta t) u(t) + \xi(\Delta t) \ , 
\end{equation}
where $\Gamma$ is the matrix associated with the parameters of the autoregressive model, we refer it as memory term, 
and $\xi$ is a white noise characterized by the Gaussian distribution
$\mathcal{N}(0,\tilde{\Sigma}_{u|_\MM})$ with  $\tilde{\Sigma}_{u|_\MM}$ the covariance of the noise given a certain parametrization $\MM$
of the AR process. If each singular value of $\Gamma$ is smaller than $1$, the process is stationary,
a state that the flow at the CMB is certainly fulfilling. Under such conditions, the time-averaged spatial covariance of the velocity 
field $\Sigma_{u|_\MM}$ is related to the covariance of the white noise $\xi$ through the relation:
\begin{equation}
 \tilde{\Sigma}_{u|_\MM} = \Sigma_{u|_\MM} - \Gamma \Sigma_{u|_\MM} \Gamma^T \ .
\end{equation}
Defining the vector ${\bf u}= \left(u_0,u_1,\cdots,u_{N-2},u_{N-1}\right)^T$ containing the velocity fields at $N$
different epochs, its full covariance ${\bf \Sigma_{u|_\MM}}$ is given by:
\begin{equation}
 {\bf \Sigma_{u|_\MM}} =
 \begin{pmatrix}
  \Sigma_{u|_\MM} & \Gamma\Sigma_{u|_\MM} & \cdots & \Gamma^{(N-1)}\Sigma_{u|_\MM} \\
  \Sigma_{u|_\MM}\Gamma^T & \Sigma_{u|_\MM} & \cdots &  \Gamma^{(N-2)}\Sigma_{u|_\MM} \\
  \vdots & \vdots & \ddots & \vdots  \\
  \Sigma_{u|_\MM}{\Gamma^{(N-1)}}^T & \Sigma_{u|_\MM}{\Gamma^{(N-2)}}^T & \cdots & \Sigma_{u|_\MM}
 \end{pmatrix}\ .
\end{equation}

\noindent If  $\Gamma$ is a scalar, this parametrization of the flow spatio-temporal
behavior is similar to the one used by \citet{Gillet2015}. Indeed, in their study they chose a temporal correlation
of the velocity field depending on the characteristic time $\tau_u$ such as $\Gamma(\Delta t) = \exp(-\frac{\Delta t}{\tau_u})$, with
$\Delta t$ the time stepping of the auto regressive process.

\subsection{Parametrization of the auto regressive process}\label{flowParameters}

\noindent To fully describe the autoregressive process, its two main parameters, the spatial covariance $\Sigma_{u|_\MM}$ and the memory term of the
process $\Gamma$, have to be characterized. Following the developments of \citet{Baerenzung2016}, $\Sigma_{u|_\MM}$ is chosen to derive from the poloidal and 
toroidal stationary spectra of the flow, the latter being assumed to behave as power laws with different spectral ranges. $\Gamma$, 
which contains the information on the temporal correlations of the velocity field, is
assumed to be scale dependent, therefore we choose it to derive from power laws presenting the same ranges than the flow energy spectra.
Under such assumptions, the poloidal and toroidal stationary spectra of the flow and memory terms of the AR process are given by:
\begin{eqnarray}
 E_\phi(l) &= C_{E_\phi}^i A_{E_\phi}^2 l^{\mbox{-}P_{E_\phi}^i}  &\quad \text{for} \quad  l \in \Delta_{\phi}^i \label{Ephi}\\
 E_\psi(l) &= C_{E_\psi}^j A_{E_\psi}^2 l^{\mbox{-}P_{E_\psi}^j}  &\quad \text{for} \quad  l \in \Delta_{\psi}^j \label{Epsi} \\
 \Gamma_\phi(l) &= C_{\Gamma_\phi}^i A_{\Gamma_\phi}^2 l^{\mbox{-}P_{\Gamma_\phi}^i}  &\quad \text{for} \quad  l \in \Delta_{\phi}^i \label{Gphi}\\
 \Gamma_\psi(l) &= C_{\Gamma_\psi}^j A_{\Gamma_\psi}^2 l^{\mbox{-}P_{\Gamma_\psi}^j}  &\quad \text{for} \quad  l \in \Delta_{\psi}^j \label{Gpsi}
\end{eqnarray}
where the $A$'s are the magnitudes of the energy spectra and the memory terms, and the $P^k$'s are their slopes 
within the spherical harmonics ranges $\Delta^k$'s. The constants $C^k$'s are given by:
\begin{equation}
 C^k =  \prod_{a=2}^{a=k} \exp\left( \log\left(l_{a\mbox{-}1}\right)\left(P^a-P^{a\mbox{-}1}\right)\right) 
\end{equation}
$l_{a}$ being the spherical harmonics degrees where transitions in slope occur. 

\noindent According to this parametrization, whereas the matrix $\Gamma$ is diagonal and contains both $\Gamma_\phi(l)$ and $\Gamma_\psi(l)$,
the spatial covariance $\Sigma_{u|_\MM}$ derives from the poloidal and toroidal energy spectra such as:
\begin{eqnarray}
\overline{\phi_{l,m} \phi_{l^\prime,m^\prime}} &=& \frac{E_\phi(l)}{l(l+1)}\delta_{l l^{\prime}} 
\delta_{m m^{\prime}} \label{covPhiPhi}\\
\overline{\psi_{l,m} \psi_{l^\prime,m^\prime}} &=& \frac{E_\psi(l)}{l(l+1))} \delta_{l l^{\prime}} 
\delta_{m m^{\prime}} \label{covPsiPsi}\\
\overline{\phi_{l,m} \psi_{l^\prime,m^\prime}} &=& 0  \quad \forall l,l^{\prime},m ,m^{\prime} \label{covPhiPsi} \ .
\end{eqnarray}

\noindent The total parameters of the autoregressive process, referred as $\MM$, can be divided into two categories, one containing
the parametrization of the spatial covariances, namely $\MM_\Sigma$, and the other associated with the memory terms of the process $\MM_\Gamma$. $\MM_\Sigma$
and $\MM_\Gamma$ are respectively given by:
\begin{eqnarray}
 \mathcal{M}_\Sigma &=& \left\{A_{E_\phi},P_{{E_\phi}}^i,\Delta_{\phi}^i,
A_{E_\psi},P_{{E_\phi}}^j,\Delta_{\psi}^j\right\} \label{parametersSigma} \\ 
 \mathcal{M}_\Gamma &=& \left\{A_{\Gamma_\phi},P_{{\Gamma_\phi}}^i,\Delta_{\phi}^i,
A_{\Gamma_\psi},P_{{\Gamma_\psi}}^j,\Delta_{\psi}^j\right\} \label{parametersSigma} \ . 
\end{eqnarray}
\noindent  where the index $i$ and $j$ are associated with the different poloidal and toroidal spherical harmonics ranges.
Once $\MM = \left\{\mathcal{M}_\Sigma, \mathcal{M}_\Gamma \right\}$ is given, the prior distribution of the velocity field can be expressed and reads:
\begin{equation}
 p({\bf u}|\MM) = {\bf \mathcal{N}}({\bf 0},{\bf \Sigma_{u|_\MM}}) \label{priorU} \ .
\end{equation}

\subsection{Posterior distribution the autoregressive parameters}

\noindent To forecast the evolution of the velocity field in the ensemble Kalman filter algorithm,
the parameters of the autoregressive process have to be known. Their posterior distribution, $p(\MM|{\bf \bar{\gamma}^o})$,  can
be expressed through $p({\bf u},{\bf b},\MM|{\bf \bar{\gamma}^o})$, the joint posterior distribution of the AR parameters, the core mantle boundary 
velocity field and magnetic field, such as:
\begin{eqnarray}
 p(\MM|{\bf \bar{\gamma}^o})  &=& \iint p({\bf u},{\bf b},\MM|{\bf \bar{\gamma}^o})\mathrm{d}{\bf u}  \mathrm{d}{\bf b} \\
 &=& \frac{1}{p({\bf \bar{\gamma}^o})}\iint p({\bf \bar{\gamma}^o}|{\bf u},{\bf b},\MM) p({\bf b}) p({\bf u}|\MM)p(\MM)\label{marginalParameters} 
 \mathrm{d}{\bf u}  \mathrm{d}{\bf b} \nonumber\ ,
\end{eqnarray}
where ${\bf \bar{\gamma}^o}$ is the observed secular variation. For this study, ${\bf \bar{\gamma}^o}$ is taken from the COV-OBS.x1 model
of \citet{Gillet2013,Gillet2015b}, and the covariance matrix of the secular variation ${\bf \Sigma_{\gamma}^o}$, is derived
from the $100$ ensemble members provided by the model. Because of the singular nature of the covariance matrix, only its diagonal part is kept.
Under such conditions, the likelihood distribution reads:
\begin{equation}
 p({\bf \bar{\gamma}^o}|{\bf u},{\bf b},\MM) = \mathcal{N}(-{\bf A} ({\bf ub}),{\bf \Sigma_{\gamma}^o}) \ .
\end{equation}

\noindent The prior distribution of the magnetic field $p({\bf b})$ is decomposed into two parts. The first part provides
the statistical properties of the large scale field ${\bf b}^<$, whereas the second part describes our prior knowledge on the small scale
field ${\bf b}^>$. The large scale magnetic field is characterized by the prior distribution:
\begin{equation}
 p({\bf b}^<) = \mathcal{N}({\bf \bar{b}^o},{\bf \Sigma_{b}^o}) \ ,
\end{equation}
where ${\bf \bar{b}^o}$ and ${\bf \Sigma_{b}^o}$ are respectively the COV-OBS.x1 magnetic field and covariance matrix. As for the 
secular variation, ${\bf \Sigma_{b}^o}$ is evaluated with the $100$ ensemble members of the model, and only its diagonal part of kept.

\noindent The small scale magnetic field is chosen to be at any time isotropically distributed, with a $0$ mean and a covariance ${\bf \Sigma_{b^>}}$
deriving from the extrapolation of the large scale field spectrum $E_{b^<}(l)$. Here we chose the formulation proposed by \citet{Buffett2007} to 
characterize the magnetic field spectrum at the CMB. It reads:
\begin{equation}
E_{b^>}(l) = C_1 \chi^l \label{buffettExt} 
\end{equation}
where $\chi=0.99$. To determine the constant $C_1$, we used the COV-OBS.x1 magnetic field sampled every two years between
$1900.0$ and $2014.0$, and performed a weighted least square fit of the associated energy spectra between SH degree $l=2$ and $l=13$.
We obtained that $C_1 = 7.15 \times 10^9$ nT${}^2$. Note that another type of extrapolation has also been tried, assuming an exponential decay
of the magnetic field spectrum. Although we do not show the results associated with this modeling, we observed that such an assumption would
provide insufficient levels of energy at small scales, leading to suboptimal predictions of the magnetic field evolution.
From the extrapolation given in equation (\ref{buffettExt}) we construct the covariance of the small scale magnetic field $\Sigma_{b^>}$ at a given time through
the relation:
\begin{equation}
 \overline{b_{l,m}^> b_{l^\prime,m^\prime}^>} = \frac{ E_{b^>}(l)}{(l+1)(2l+1)}  \delta_{l l^{\prime}} 
\delta_{m m^{\prime}} \ .\label{SSBcov}
\end{equation}
Neglecting a priori the temporal correlations between the small scales of the magnetic field, the full covariance ${\bf \Sigma_{b^>}}$ is simply
a bloc diagonal matrix where every block are identical and given by $\Sigma_{b^>}$.

\noindent As expressed in the previous section, the prior distribution of the velocity field conditioned by the AR parameters $p({\bf u}|\MM)$ 
is given by equation (\ref{priorU}).

\noindent The last distribution entering equation (\ref{marginalParameters}), is the prior distribution of the AR parameters $p(\MM)$. These parameters
are depending on the magnitudes $A$'s, the slopes ${P}$'s, and the spherical harmonics ranges ${\Delta}$'s of the flow stationary spectra and the AR memory terms.
Whereas the ranges $\Delta$'s will be a priori imposed, and therefore considered as known, the magnitudes and slopes are completely undetermined. 
To reflect this lack of knowledge we characterize them by uniform distributions such as:
\begin{eqnarray}
 p(A) &=& \mathcal{U}(0, \infty) \label{priorMA} \\
 p(P) &=& \mathcal{U}(-\infty, \infty) \label{priorMP} \ .
\end{eqnarray}
The full prior distribution of $\MM$ is simply the product of the prior distributions of each individual AR parameter.

\noindent Finally, following the development of \citet{Baerenzung2016}, the posterior distribution of the AR parameters given the secular variation
 of equation (\ref{marginalParameters}) is approximated by the following distribution:
\begin{equation}
 p(\MM|{\bf \bar{\gamma}^o}) = \frac{\exp\left[-\frac{1}{2} { \mathbf{ \bar{\gamma}} ^{\mathbf{o}T}  } \bf{\Sigma_{\MM_{|\bar{\gamma}^o}}^{-1}} \bf{\bar{\gamma}^o}\right]}
 {(2\pi)^{\frac{d}{2}}|\bf{\Sigma_{\MM_{|\bar{\gamma}^o}}}|^{\frac{1}{2}}} \label{posteriorParameter}
\end{equation}
\noindent where $d$ is the dimension of the secular variation vector.
To construct the matrix $\bf{\Sigma_{\MM_{|\bar{\gamma}^o}}}$, the covariance between ${\bar{\gamma}^o}$ at a time $t_\alpha$
and ${\bar{\gamma}^o}$ at a time $t_\beta$, with respect to the distribution $p({\bf \bar{\gamma}^o}|{\bf u},{\bf b},\MM) p({\bf b}) p({\bf u}|\MM)$
is calculated for every combination of epochs considered. The component at a row index $i$ and a column index $j$ of the 
resulting covariance matrix $\Sigma_{\MM_{|\bar{\gamma}^o}}^{t_\alpha t_\beta}$reads:
\begin{eqnarray}
 \left(\Sigma_{\MM_{|\bar{\gamma}^o}}^{t_\alpha t_\beta} \right)_{ij} & = & \left(\Sigma_{\gamma}^{o^{t_\alpha t_\beta}}\right)_{ij} + 
 \left( A_{\bar{b}_{o}}(t_\alpha)\Sigma_{u_{|\MM}}^{t_\alpha t_\beta}  A_{\bar{b}_{o}}^T(t_\beta)\right)_{ij} \nonumber\\
  & & + A_{imn} \left(\Sigma_{u_{|\MM}}^{t_\alpha t_\beta}\right)_{mr}
 \left(\Sigma_{b}^{o^{t_\alpha t_\beta}}\right)_{ns} A_{jrs}  \ ,
\end{eqnarray}
where we recall that the third order tensor $A$ is defined such as $A_{ijk}(u)_j(b)_k = \left(A_{b} u \right)_i = \left(A_u b \right)_i$.

\subsection{Sequential assimilation of the core secular variation and magnetic field}

\noindent To combine our dynamical model for the magnetic field and the velocity field at the CMB to
a magnetic field and secular variation model derived from geomagnetic data, we implemented
the Ensemble Kalman filter approach proposed by \citet{Evensen2003}.
This method proceeds in two steps. In the first one, referred as the prediction step, the spatio temporal evolution of the 
state variables (the velocity and magnetic fields) represented through an ensemble is forecasted until data become available. 
Then, the second step, called the analysis, is initiated, and
each member of the ensemble is corrected accordingly to the data.

\noindent As mentioned previously, while the dynamical behavior of the velocity field is prescribed by a first order auto regressive process,
the evolution of the magnetic field is constrained by the frozen flux equation. Nevertheless, directly solving the FF equation is numerically
unstable. Indeed, since in this equation has no diffusion mechanism, cascading magnetic energy will have a tendency
to accumulate on the smallest simulated scales, and slowly contaminate the entire field through non linear interactions. 
To counter this latter effect an extra hyperdiffusion term is added to the FF equation. Under such conditions the prediction step of the EnKF 
for each member $k$ of the  ensemble of velocity and magnetic fields $\left\{u,b \right\}$ is given by:
\begin{eqnarray}
 u_k(t+\Delta_t) &=& \Gamma(\Delta_t)u_k(t) + \xi_k(\Delta_t) \label{forecastU}\\
 \partial_t b_k(t) &=& -A({u_k}(t) b_k(t)) - \eta_D \Delta_H^4 b_k(t)\label{forecastB}
\end{eqnarray}
 where the memory term $\Gamma$ and the random noise $\xi$ of the AR process are scaled accordingly to the time step $\Delta t$.
 The hypperdiffusivity is set to $\eta_D = 9\times 10^{13}$ km${}^8$.yr${}^{-1}$. If the magnetic field was only diffused over time with 
 such an hyperdiffusion, in $100$ years the latter would loose $0.09\%$, $18\%$ and $99\%$ of energy at respectively spherical harmonics degrees
 $13$, $26$ and $39$.
 The numerical resolution of the FF equation is performed through an Euler scheme for the first iteration, and a second order Adams-Bashforth
 scheme for the following ones. The time step $\Delta_t$ has been set to half a year.
 
 \noindent For the analysis step of the EnKF,
 both the observed magnetic field and secular variation are assimilated simultaneously. To do so, the model state has to be 
 augmented in order to take the secular variation into account. Therefore, from each pair $(u_k^f,b_k^f)$ of the forecasted ensemble, 
 a prediction for the observables is build accordingly to the following relations:
\begin{eqnarray}
 \gamma_k^f &=& -A(u_k^f b_k^f) \label{gammaFor}\\
 b^{f<}_k &=& H b_k
\end{eqnarray}
where the linear operator $H$ simply truncates the forecasted magnetic field at the level of the observed one.
Normally equation (\ref{gammaFor}) should contain the hyperdiffusion term of equation (\ref{forecastB}). However, the effects of this hyperdiffusion
are so weak on the range of scales describing the observed secular variation (the spectral expansion of the 
COV-OBS.x1 SV does not excess SH degree $l=13$ in this study), that they are neglected.

\noindent From the augmented ensemble $\left\{u^f,b^f,\gamma^f,Hb^f \right\}$ the covariances necessary for the analysis step of the EnKF are calculated.
Recalling that the covariance of a field $a^f$ is referred as $\Sigma_a^f$ and the covariance between a field $a^f$ and a field $b^f$ is expressed 
as $\Sigma_{ab}^f$, each updated pair of velocity field and magnetic field $(u_k^a,b_k^a)$ is given by:
\begin{eqnarray}
 \begin{pmatrix}
      u_k^a \\
      b_k^a 
 \end{pmatrix}
 &=&
 \begin{pmatrix}
      u_k^f \\
      b_k^f 
 \end{pmatrix}
  + 
 \begin{pmatrix}
  \Sigma_{u\gamma}^f & \Sigma_{ub}^fH^T \\
  \Sigma_{b\gamma}^f & \Sigma_{b}^f H^T
  \end{pmatrix}\\ 
  &\times& 
 \begin{pmatrix}
  \Sigma_{\gamma}^f+\Sigma_{\gamma}^o & \Sigma_{\gamma b}^fH^T \\
  H\Sigma_{b\gamma}^f & H\Sigma_{b}^f H^T + \Sigma_{b}^o
  \end{pmatrix}^{-1} 
 \begin{pmatrix}
 \gamma_{k}^o - \gamma_k^f\\
 b_{k}^o - Hb_k^f
 \end{pmatrix}\nonumber \label{enfk}
\end{eqnarray}
where $\gamma_{k}^o$ and $b_{k}^o$ are random realizations from the distributions of the COV-OBS.x1 model for respectively 
the secular variation $p(\gamma^0)=\mathcal{N}(\bar{\gamma}^o,\Sigma_{\gamma^o})$ and the magnetic field  $p(b^0)=\mathcal{N}(\bar{b}^o,\Sigma_{b^o})$.

\section{Geophysical application}\label{results}
\subsection{Numerical set up}\label{numericalSetUp}

\noindent The time period considered in this study is $1900.0-2014.0$, and the magnetic field and secular variation data are taken from the COV-OBS.x1 model
with a $2$ year sampling rate, corresponding to the knots of the model's B-spline expansion.
Every simulation is performed through a pseudo spectral approach on the Gauss-Legendre grid provided by \citet{Schaeffer2013}.
Both the poloidal and toroidal parts of the velocity field are expanded up to SH degree $l=26$, and the radial component
of the magnetic field is expressed up to SH degree $l=39$ in order for the field to possess a large enough diffusion range.
Whereas the COV-OBS.x1 magnetic field is always taken up to SH degree $l=13$, the expansion of the COV-OBS.x1 secular variation depends on the variance 
level associated with each scale. If globally at a certain scale the standard deviation of the secular variation is larger than the absolute value of the
mean field, the total field is truncated at this scale. Under such a condition, the COV-OBS.x1 secular variation is taken up SH degrees $l=10$, $l=11$, $l=12$ and
$l=13$ for the respective time windows $[1900-1923]$, $[1924-1943]$, $[1944-1963]$ and $[1964-2010]$.
Finally, the state of the system is characterized by $40000$ pairs of magnetic field and velocity field at the core mantle boundary.

\subsection{Estimation of the flow optimal auto regressive parameters}\label{AREstimation}
\noindent To simulate the spatio temporal evolution of the flow at the CMB, the parameters of the auto regressive process have to
be estimated. We recall that their posterior distribution, $p(\MM|{\bf \bar{\gamma}^o})$,
is expressed in equation (\ref{posteriorParameter}). By maximizing this distribution, one should get the optimal
parameters for the AR process. However, instead of estimating both temporal and spatial parameters simultaneously,
we proceed in two steps. 
First, only the spatial
covariance of the velocity field is evaluated following the method proposed and tested by \citet{Baerenzung2016}. 
This approach consists in maximizing the distribution $p(\MM|{\bf \bar{\gamma}^o})$ in which only the block diagonal
part of the covariance matrix $\Sigma_{\MM_{|\bar{\gamma}^o}}$ is kept. 
Once the spatial covariance is determined, it is assumed to be known, and the maximum of $p(\MM_\Gamma|{\bf \bar{\gamma}^o},\hat{\MM}_\Sigma)$ is calculated.

\noindent Following the developments of section \ref{flowParameters}, the AR parameters are decomposed into scale dependent power laws 
exhibiting different spectral ranges. \citet{Baerenzung2016} showed that if the stationary spectra of the flow are decomposed into two
spectral ranges, the optimal scales where transition in slope occurs are $l=3$ and $l=8$ for respectively the toroidal and poloidal energy
spectra. Here, whereas we keep the same decomposition for the spectrum associated with the poloidal field, more degrees of freedom are allowed 
for the toroidal field spectrum. Since we wish to accurately determine the spatio-temporal evolution of the eccentric gyre, toroidal field component at
SH degree $l=1$ and $l=2$, the main components of the gyre, are free to exhibit any variance level and characteristic time. Similarly, toroidal 
field components at SH degree $l=3$ are also assumed to be unconstrained by surrounding velocity field scales. This choice is motivated by
the particular low level of energy that these scales are exhibiting over recent epochs (see \citet{Baerenzung2016, Whaler2016}). 
Finally, one spectral range is used to characterize the toroidal 
field spatial variance and memory effects between SH degrees $l=4$ and $l=26$. So all in all, the AR parameters associated with the toroidal field exhibit 
the four respective spectral ranges, $\Delta_0 = [1]$, $\Delta_1 = [2]$, $\Delta_2 = [3]$ and $\Delta_3 = [4,26]$.

\begin{figure}[h]
\begin{center}
      \includegraphics[width=0.95\linewidth]{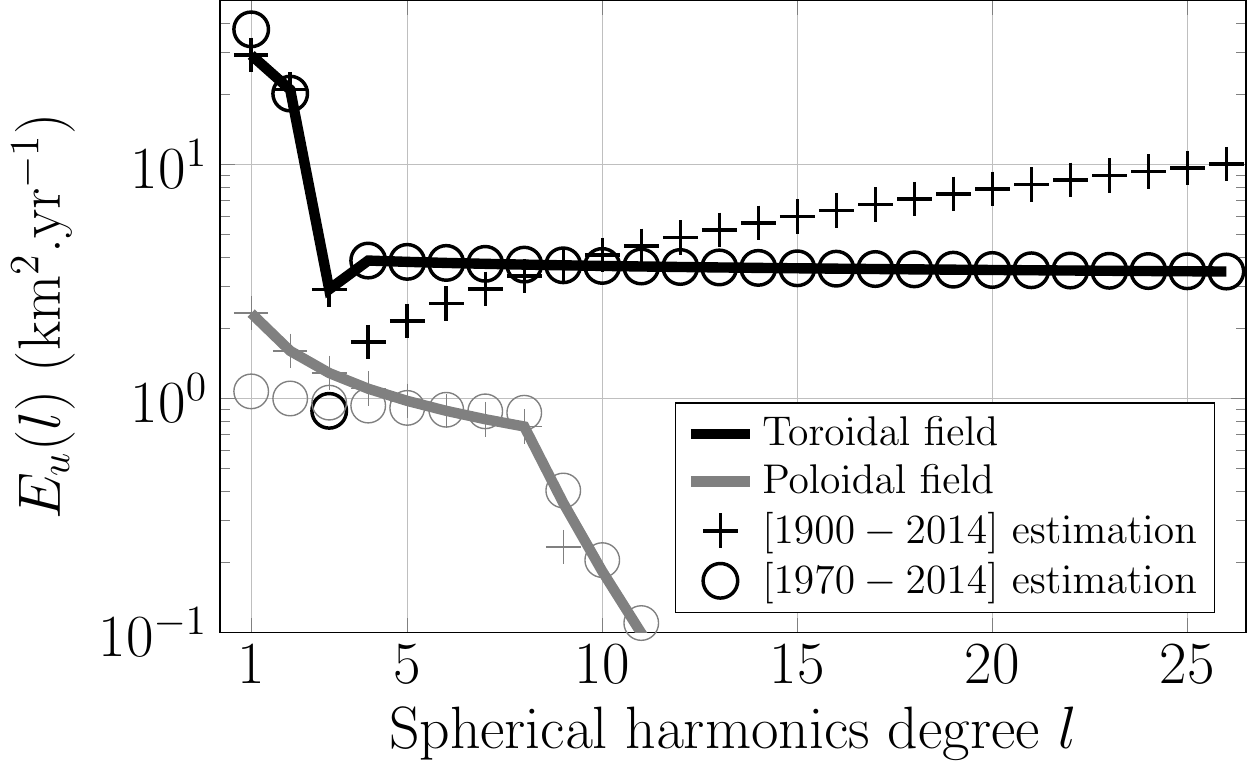}
\caption{Prior kinetic energy spectra for the toroidal part of the velocity field (black) and for its poloidal part (gray). 
Estimations with the COV-OBS.x1 secular variation and magnetic field model between $1900.0$ and $2014.0$ (crosses) and between
$1970.0$ and $2014$ (circles). The solid lines are the combination of the two evaluations used as a prior information to parametrize
the autoregressive process for the flow.}\label{priorSpectra}
\end{center}
\end{figure}

\noindent As mentioned in the beginning of the section, the estimation of the stationary energy spectra parameters is performed between $1900.0$ and $2014.0$,
taking the COV-OBS.x1 magnetic field and secular variation every $\Delta_t=2$ years. On figure \ref{priorSpectra} the resulting power law spectra are displayed
with crosses. As already observed in \citet{Baerenzung2016}, the toroidal field (in black), and in particular its large scales (SH degree $l=1$ and $l=2$), 
exhibits a much larger energetic level than the poloidal field (in gray). Nevertheless the toroidal energy spectrum also presents a strong increase of energy
towards its smallest scales.
This effect, which is in contradiction with the results of \citet{Baerenzung2016}, is very likely to be attributed to a slight underestimation 
of the COV-OBS.x1 secular variation uncertainties. 
Therefore, in order to better estimate the small scale energy spectra of the velocity field, we performed different estimations of the stationary spectra
parameters by varying the time window in which the evaluation is computed. We found that the largest period where the spectra did not exhibit an anomalous behavior was 
$1970.0-2014.0$. The resulting prior spectra are shown in figure \ref{priorSpectra} with circles. Combining the small scale spectra of the $1970.0-2014.0$ evaluation
to the large scale ones of the $1900.0-2014.0$ estimation, we get the final prior spectra for both the toroidal and poloidal field displayed
in figure \ref{priorSpectra} with solid lines. The values associated with the spectra parameters are given in table \ref{spectraTable}.

\begin{table}
\caption{Combined optimal covariance parameters $\hat{\MM}_\Sigma$ of the flow auto regressive process,
within the $1970.0-2014.0$ and the  $1900.0-2014.0$ periods.  $A_i$ correspond to the magnitudes and $P_i$ to the slopes 
of the prior stationary spectra, within the spectral ranges $\Delta_i$
(see equations (20-21) and (25-26)).}
\centering
\begin{tabular}{c c c c c }
\hline
 Flow field & index $i$ & $\Delta_i$ & $A_{i}$ & $P_i$   \\
\hline
\multirow{4}{*}{Toroidal } & $1$ & $1$ & $5.41$ & $0$   \\
                               & $2$ & $2$ & $4.56$ & $0$   \\
                               & $3$ & $3$ & $1.71$ & $0$   \\
                               & $4$ & $[4\ ,26]$ & $2.05$ & $5.8\times 10^{-2}$  \\
\hline
\multirow{2}{*}{Poloidal } & $1$ & $[1\ ,8]$ & $1.52$ & $0.54$   \\
                               & $2$ & $[8\ ,26]$ & $663$ & $6.4$ \\
\hline
\label{spectraTable}
\end{tabular}
\end{table}

\begin{figure}[h]
\begin{center}
      \includegraphics[width=0.95\linewidth]{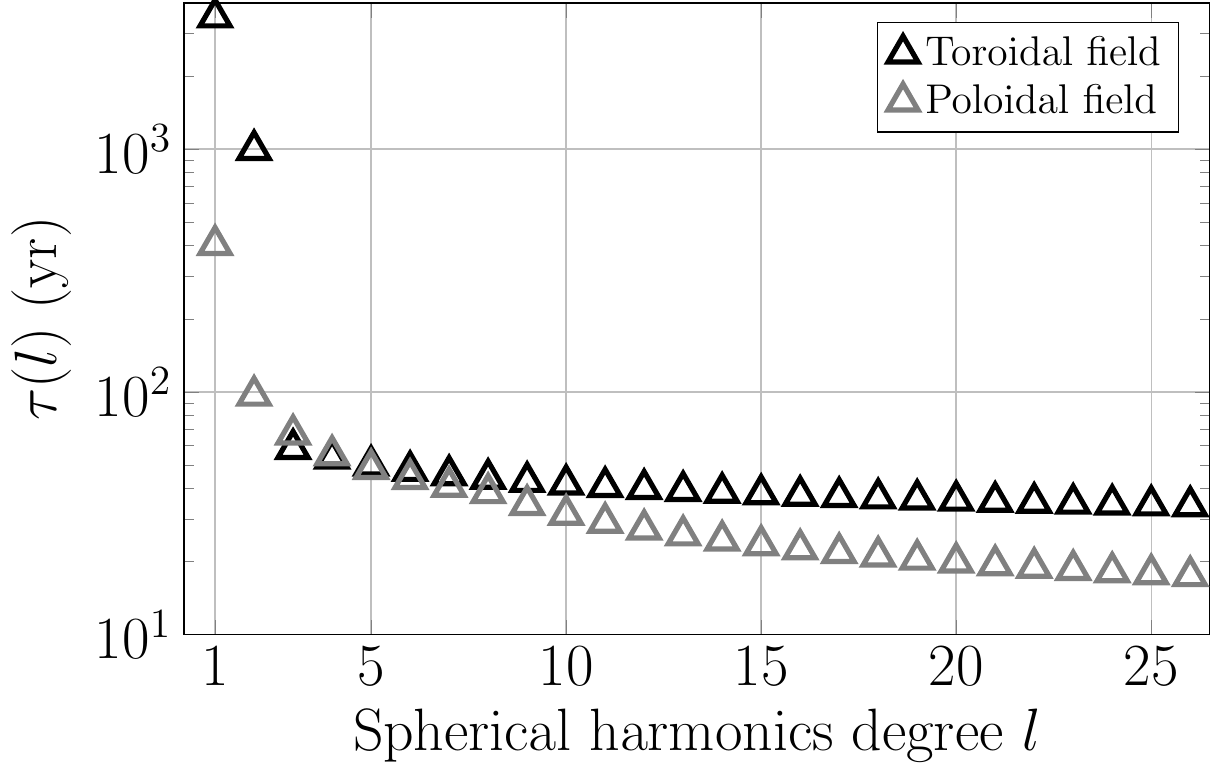}
\caption{Prior characteristic timescale $\tau(l)$ for the autoregressive process of the flow, associated with the toroidal part of the velocity field
(in black) and its poloidal part (in gray).
}\label{timescale}
\end{center}
\end{figure}

\noindent The spatial covariances of the AR process being characterized, the evaluation of the memory terms is now performed by maximizing the distribution 
$p(\MM_\Gamma|{\bf \bar{\gamma}^o},\hat{\MM}_\Sigma)$  within the $1900.0-2014.0$ time window. The results, expressed through the scale dependent characteristic
time $\tau(l) = -\frac{\Delta_t}{\log(\Gamma(l))}$ are shown on figure \ref{timescale} and the parameters of the memory terms are given in table \ref{tableParameters}. 
The most striking feature one can observe, is the very long
memory time (of the order of thousand years) associated with the main component of the eccentric gyre. 
This indicates that this structure is very persistent over time. Nevertheless, these values should be taken with care since
their evaluation is performed on a comparatively short time window of $114$ years. In contrast with the large scale field, the toroidal field components at 
spherical harmonics degree varying from $l=3$ to $l=26$ exhibit lower characteristic memory times, with a decaying 
behavior where $\tau(l=3)\sim 50$ yr and $\tau(l=26)\sim 30$ yr.
Note that this limiting time is similar to the e-folding time of the Geodynamo as calculated by \citet{Hulot2010,Lhuillier2011}.
The characteristic times associated with the poloidal field indicate that the latter presents a slower dynamical behavior at large scales than at
small scales with $\tau(l)$ varying from $\tau(l=1)\sim 400$ yr to $\tau(l=8) \sim 40$ yr.

\begin{table}
\caption{Optimal parameters for the memory term of the auto regressive process $\hat{\MM}_\Gamma$,
evaluated within the $1900-2014$ time window.  $A_i$ and $P_i$  are respectively the magnitudes and the slopes 
of the assumed power laws within the spectral ranges $\Delta_i$. Also given are the characteristic times of the 
AR process $\tau = -\frac{1}{\log(\Gamma)}$ expressed in years. }
\centering
\begin{tabular}{c c c c c c c}
\hline
Flow field   & index $i$ & $\Delta_i$ &$1-A_{i}$ & $P_i$ &$\tau(\Delta_i)$  \\
\hline
\multirow{4}{*}{Toroidal }        & $1$ & $1$  &  $1.43\times 10^{-4}$ & $0$ & $3495$  \\
                                  & $2$ & $2$  & $5.03\times 10^{-4}$  & $0$ & $994$  \\
                                  & $3$ & $3$  & $8.53\times 10^{-3}$  & $0$ & $58$  \\
                                  & $4$ & $[4\ ,26]$ & $5.42\times 10^{-3}$ &  $5.72\times 10^{-3}$  & $[53\ ,34]$  \\
\cline{1-6}
\multirow{2}{*}{Poloidal }        & $1$ & $[1\ ,8]$ & $1.24\times 10^{-3}$   &  $1.13\times 10^{-3}$ & $[403\ ,38]$  \\
                                  & $2$ & $[8\ ,26]$ & $-1.47\times 10^{-2}$  &  $2.65\times 10^{-2}$ & $[38\ ,17]$ \\
\hline
\label{tableParameters}
\end{tabular}
\end{table}

\subsection{General properties of the flow at the core mantle boundary}\label{flow}

\noindent The AR process for the flow being parametrized, estimation of the velocity field and magnetic field at the CMB 
through the EnKF algorithm can be performed. To initialize the fields in $1900.0$, we applied the Gibbs sampling algorithm
proposed by \citet{Baerenzung2016} and which is detailed in the appendix \ref{appA}. However, instead of sampling the joint posterior distribution of the flow and the magnetic field
at the $1900.0$ epoch only, we sampled the distribution characterizing simultaneously the fields in $1900.0$, $1950.0$ and $2000.0$, 
in order to constraint the initial state with recent observations.

\begin{figure}[h]
\begin{center}
      \includegraphics[width=0.95\linewidth]{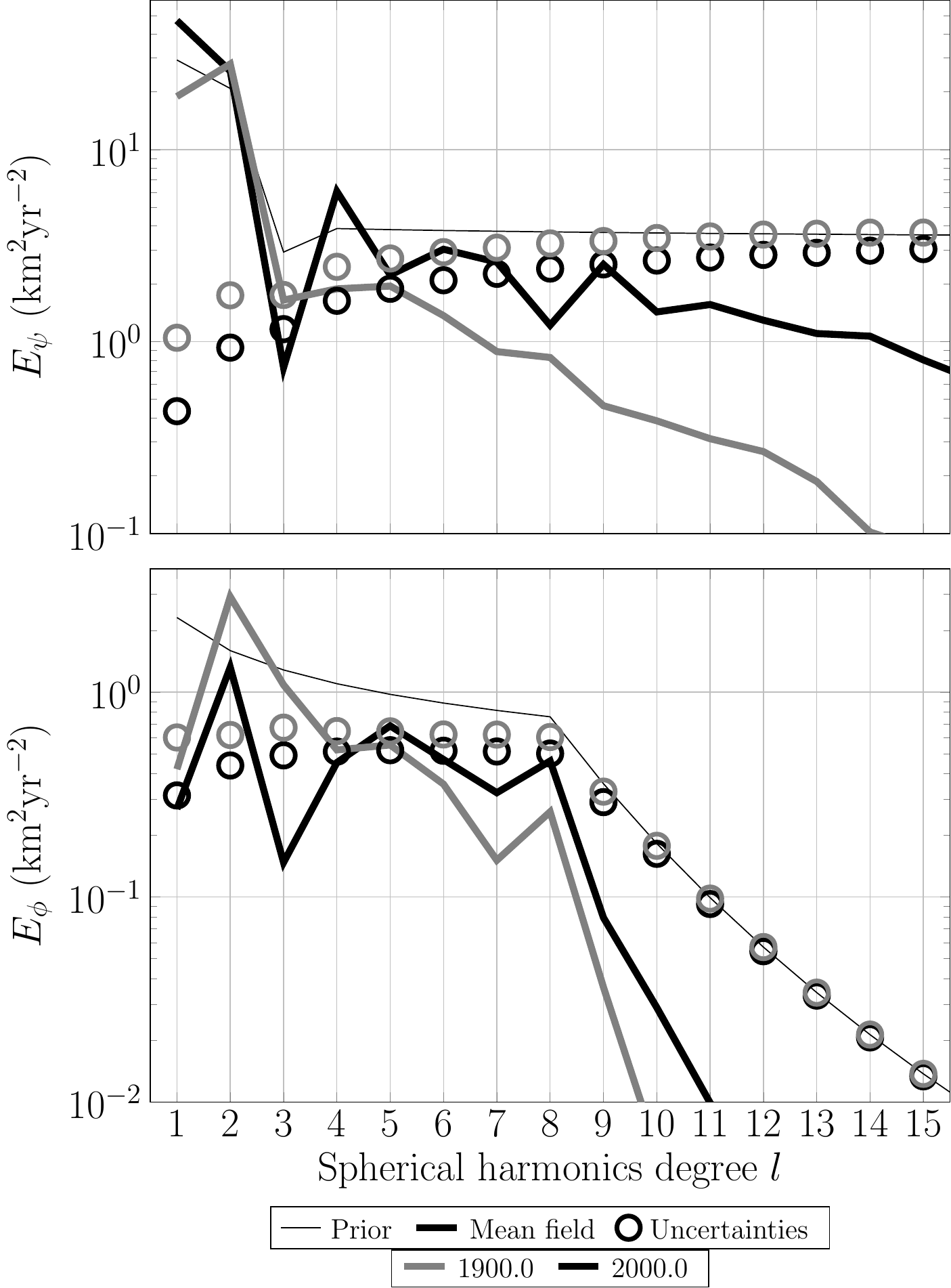}
\caption{Toroidal (top) and poloidal (bottom) energy spectra associated with the ensemble mean fields (thick lines) and standard deviation
(circles), for the $1900.0$ (gray) and $2014.0$ (black) epochs. Thin lines correspond to the prior spectra. }\label{spectra}
\end{center}
\end{figure}

\noindent Because of the sequential nature of the EnKF algorithm, accuracy of the estimated fields is not constant over time, but increases whenever new data are 
assimilated. This effect is well illustrated by figure \ref{spectra} displaying at two different epochs, $1900.0$ (grey) and $2000.0$ (black),
the energy spectra of the toroidal (top) and poloidal (bottom) mean velocity fields (thick solid lines) and uncertainty fields (circles). 
Whereas in $1900.0$ the mean toroidal 
field exhibit a level of energy larger or of the same order than the variance of the field up to SH degree $l=3$, in $2000.0$ reliable information become
available up to SH degree $l=9$. Above these scales the posterior variance of the flow rapidly reaches its prior level. One can also
notice that the larger the scale, the stronger the variance reduction over time. The latter observation, which is also valid for the poloidal field, is
linked to the behavior of the characteristic times $\tau(l)$ associated with the different flow scales (see figure \ref{timescale}). 
The fact that $\tau(l)$ is a strictly decaying function of $l$, implies that small scales velocity field will exhibit a higher randomization rate
than large scales, and so between two analysis step, the prior variance will increase faster at small than at large scales.

\begin{figure}[h]
\begin{center}
      \includegraphics[width=0.95\linewidth]{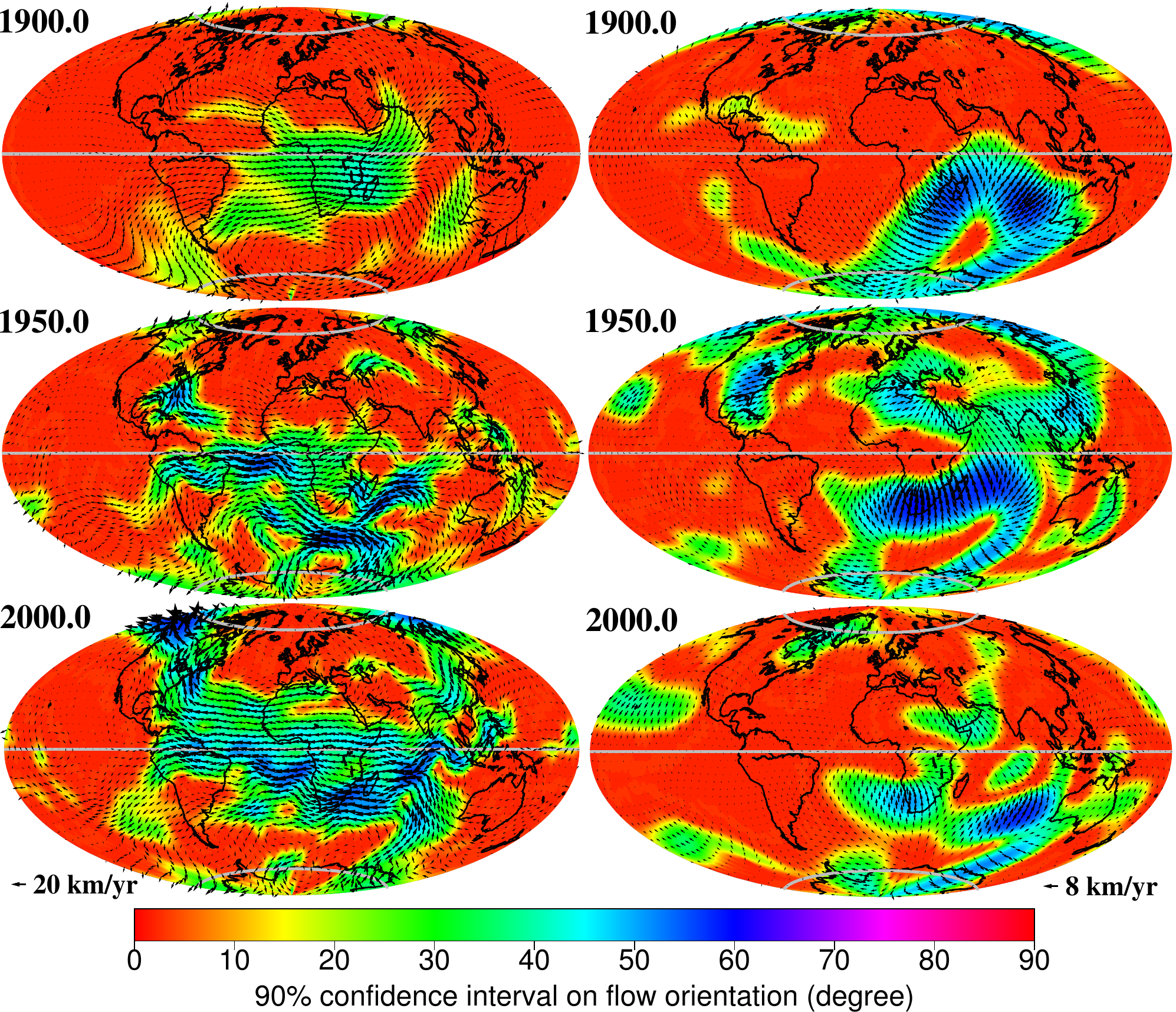}
\caption{Toroidal (left) and poloidal (right) velocity fields in $1900.0$ (top), $1950.0$ (middle) and $2000.0$ (bottom).
Arrows correspond the mean fields (over the ensemble), and color maps are displaying the $90\%$ confidence interval on their orientation.
Note that the scaling for the velocity field is different between the toroidal and poloidal part of the flow.}\label{flows}
\end{center}
\end{figure}

\noindent In physical space, the gain of flow accuracy over time is particularly stricking for the toroidal part of the velocity field as shown on figure 
\ref{flows}. In this figure, the toroidal (left) and poloidal (right) mean velocity fields 
are displayed with black arrows for three different epochs, $1900.0$ (top), $1950.0$ (middle)
and $2000.0$ (bottom). Color maps, representing the $90\%$ confidence interval on the velocity field orientation, provide information on
locations where the mean flow direction can be reliably estimated (violet and blue) or not (red). 
In $1900.0$, very little parts of the eccentric gyre can be confidently estimated. Only the  westward flow below Africa and the Atlantic ocean, 
the Southern branches of the gyre, and the Northern circulation around and partially inside the tangent cylinder (the cylinder tangent 
to the inner core and aligned with the axis of rotation of the Earth), appear as reliable patterns. 
At later times the gyre is well defined, and many of its small scale structures become visible. Globally, uncertainties on the toroidal part of the flow
are decreasing with time. 
This does not seem to be the case for the poloidal field where in $1950.0$ reliable patterns are covering a larger surface of the CMB than in $2000.0$. 
Nevertheless, the r.m.s velocity of the poloidal field and associated standard deviation 
of $3.49\pm0.23$ in $1950.0$ and $2.49\pm0.21$ in $2000.0$,
indicate that the global uncertainty level of the poloidal field as well as its magnitude have been decreasing between $1950.0$ and $2000.0$.

\noindent Contrary to flow models constrained to be geostrophic (see \citet{Bloxham1991,Amit2013}), up and downwelling fluid motions, 
which converts into poloidal field when meeting the CMB, are not particularly located around the equator. Instead, a strong and persistent 
poloidal structure evolves below the Indian ocean and South Africa. According to \citet{Bloxham1986}, such a poloidal field could be at the origin,
through the expulsion of magnetic flux from the outer core, of the intense reversed flux patch located there. Although the frozen
flux equation cannot model the transport of magnetic structures between the core and its outer boundary, 
poloidal field spreading or concentrating magnetic patches around location of intense up or downwelling motions can be detected.

\begin{figure*}[h]
\begin{center}
      \includegraphics[width=0.85\linewidth]{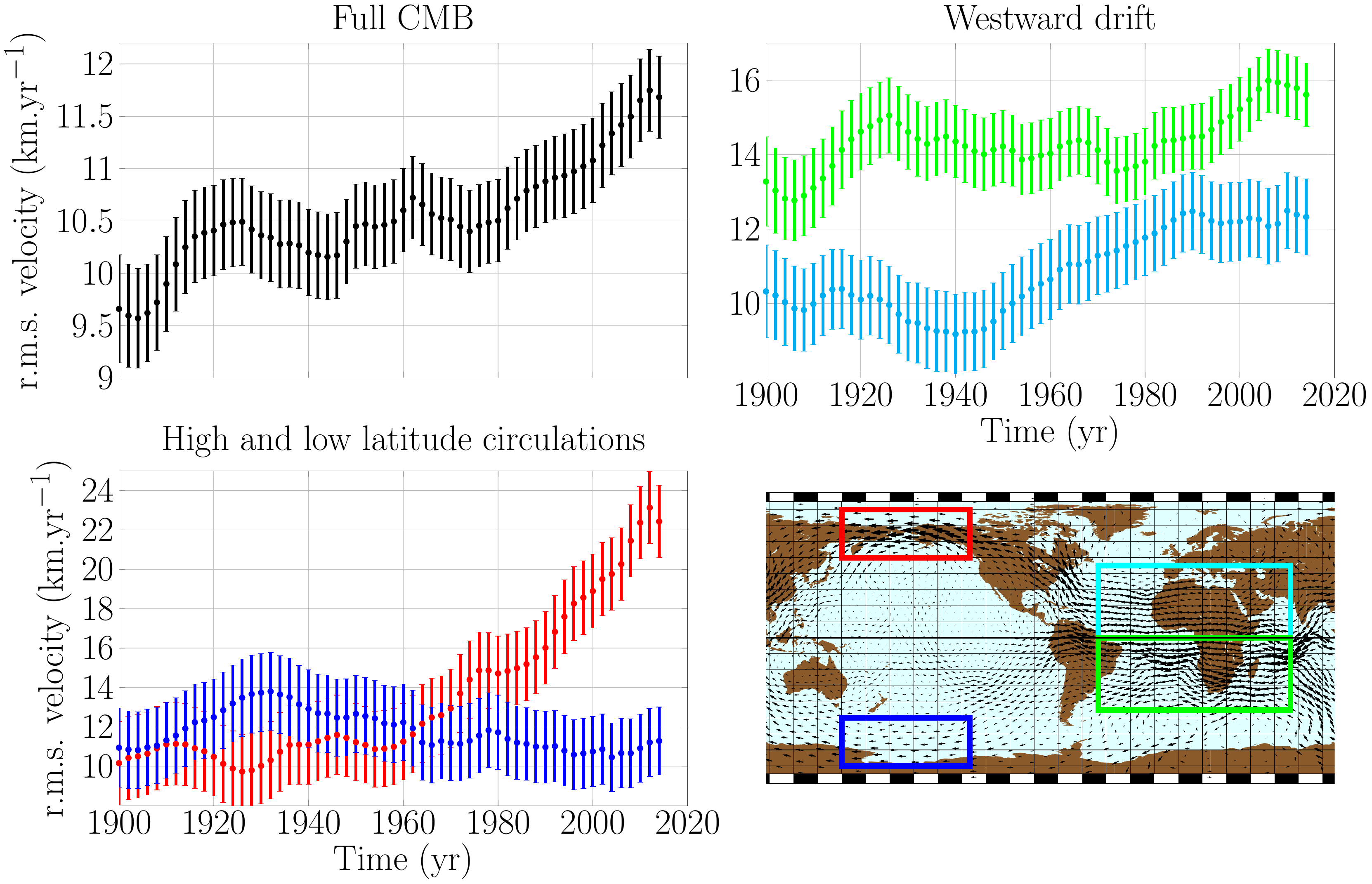}
\caption{r.m.s. velocity of the toroidal part of the flow and associated standard deviation in different locations of the core mantle boundary.
Colored curves correspond to the r.m.s. velocity within the areas surrounded by the same colored contour on the bottom right of the figure.
The black and gray symbols on the top left corresponds to the r.m.s. velocity of respectively the toroidal flow and its
equatorial symmetric part over the entire surface of the CMB. The arrows on the bottom right are associated
with the $1980.0-2014.0$ time averaged mean toroidal field.
}\label{rms}
\end{center}
\end{figure*}

\noindent Other specific features of the velocity field are in apparent contradiction with a possible geostrophic state of the outer core flow. 
This includes the reliable part of the toroidal field penetrating the tangent cylinder, or its component crossing the equator below India and South America
as already reported in other studies (see \citet{Barrois2017}). 
Particularly striking, is the apparent violation by the eccentric gyre of the equatorial symmetry condition imposed by quasi-geostrophy (see \citet{Amit2004}).
Indeed, the flow responsible for the westward drift together with the circulations around the tangent cylinder,
exhibit different levels of intensity in the Northern hemisphere than in the Southern one. These visual observations 
are confirmed in figure \ref{rms}, where the r.m.s. velocity and associated standard deviation of the toroidal part of the flow,
are measured in different locations of the CMB. One can observe on the top right of figure \ref{rms},
that the flow evolving below Africa and the Atlantic ocean within the areas shown on bottom right of the figure, is at any
time more intense in the South than in the North. For the circulations around the tangent cylinder, both southern and northern part
exhibit similar levels of energy between $1900.0$ and $1960.0$ as shown on the bottom left of figure \ref{rms}. However,
in $1960.0$ the flow below Alaska and the Eastern part of Siberia starts to accelerate, and intensifies almost continuously over
the last decades to reach a r.m.s. velocity around $23$ km.yr${}^{-1}$ in $2014.0$. This acceleration has already been observed
during the satellite era by \citet{Livermore2017}. Here we observe that it takes its origin at quite early times.
Although the toroidal part of the flow exhibits some clear deviation from geostrophy, comparisons of its r.m.s. velocity 
over the entire CMB (black symbols on the top left of figure \ref{rms}) with the r.m.s. velocity of its equatorial symmetric part (gray symbols),
shows that the latter component remains at any time dominant.

\subsection{Variations in length of day}\label{lodSection}

\noindent Many factors influence the Earth's rotation and thus the length of the day (LOD) $\Lambda$.
Variations on short timescales are mostly caused by an exchange of angular momentum between Earth's solid body and 
atmospheric and oceanic currents. \citet{Jault1988} have shown that the coupling between core 
and mantle induced decadal LOD changes. Nevertheless, modification of the Earth's oblateness ($J_2$) through the melting of glacier and ice caps
and through the global sea level rise is also thought to have a non negligible influence on the LOD over such timescales (see \citet{Munk2002}).
Over centenial and millennial timescales, two principle mechanisms are modifying the rotation rate of the Earth. The first one is the tidal friction.
By deforming the Earth's surface, tidal forces are inducing a dissipation of energy in the Earth-Moon system. As a consequence, the rotation rate of
the Earth is decreasing, and the LOD increases with a rate of $\dot{\Lambda}_{\textsc{\tiny{MOON}}} \sim 2.4$ ms/cy as estimated by \citet{Williams2016}.
The other phenomenon modifying the long term variations in LOD is the glacial isostatic adjustment (GIA). 
When accumulating on the polar caps over the last glaciation period, ice was compressing the mantle, inducing an increase of the Earth's 
oblateness. When the ice melted, the mantle tend to regain its initial shape at a rate depending on its viscosity profile. 
\citet{Peltier2015} has shown that the associated decrease in oblateness would decrease the LOD at a rate 
of $\dot{\Lambda}_{\textsc{\tiny{GIA}}} \sim -0.6$ms/cy.

\noindent The sum of the latter effects $\dot{\Lambda}_{\textsc{\tiny{MOON}}} + \dot{\Lambda}_{\textsc{\tiny{GIA}}} \sim 1.8$ms/cy explains very well
the millennial trend of increase in LOD of $1.78$ms/cy observed by \citet{Stephenson2016}. Nevertheless, if one accounts for the influence of 
the global sea level rise and ice melting over the last century (see \citet{Church2011,Hay2015}), the trend in the 
observed variations in LOD (${\Lambda}_{\textsc{\tiny{OBS}}}$) should be slightly larger than $1.8$ms/cy.
More precisely, through measurements of the Earth's oblateness with satellite laser ranging, \citet{Cheng2013} have shown that the oblateness 
was decreasing before the 1990's, but not sufficiently to be explained by the GIA alone, and increasing after. Therefore, 
the trend in ${\Lambda}_{\textsc{\tiny{OBS}}}$  should lie between $1.8$ms/cy and $2.4$ ms/cy before the 1990's and it should be larger than
$\dot{\Lambda}_{\textsc{\tiny{MOON}}} =2.4$ ms/cy after.

\noindent As highlighted by \citet{Munk2002}, these estimations are in contradiction with the trend of $1.4$ms/cy observed over the almost last $200$ years.
Therefore, either a mechanism increasing the rotation rate of the Earth is missing, or GIA effects are underestimated. 
\citet{Mitrovica2015} showed that combining a GIA model exhibiting a lower $J_2$ rate than the one of \cite{Peltier2015}, 
to lower estimates of the global sea level rise over the last century, could explain the discrepancies between observed and predicted trend in 
${\Lambda}$. Nevertheless, for this model to be compatible with the millennial trend, 
a mechanism slowing down the rotation of the Earth over very long periods of time is necessary. 
They suggested that the outer core flow estimated by \citet{Dumberry2006} could be responsible for this. This means that the oscillations
of the core angular momentum highlighted by \citet{Dumberry2006} would also be accompanied by a global decay.

\noindent So because of the uncertain nature of the trend in ${\Lambda}$ , we estimated the optimal value it should
take accordingly to our ensembles of velocity fields. To compute the variations in LOD deriving from the flow at the core mantle boundary, 
we used the formula of \citet{Jault2015}
which reads:
\begin{equation}
 {\Lambda}_{\textsc{\tiny{FLOW}}}  = 1.232(\psi_1^0 + 1.776 \psi_3^0 + 0.08 \psi_5^0 + 0.002 \psi_7^0)\ . \label{Jault}
\end{equation}

\noindent Based on the previous arguments, we decompose the LOD  into the core contribution 
${\Lambda}_{\textsc{\tiny{FLOW}}}$ and an additional long-term linear trend with rate $a$ 
modeling the effects discussed above such as:
\begin{equation}
\Lambda(t) = a t + b + \dot{\Lambda}_{\textsc{\tiny{FLOW}}} 
\end{equation}
where $b$ is related to a reference observation at $t_0$: 
$b= \Lambda_{\textsc{\tiny{OBS}}}(t_0) - a t_0$.
Instead of prescribing $a$ and $b$ we search for their optimal values accordingly to our core flow model
in a Bayesian approach. 
Assuming that $a$ and $b$ are a priori unknown
(uniform prior over infinite ranges), the posterior distribution of $a$ can be expressed as:
\begin{equation}
 p(a|{\Lambda}_{\textsc{\tiny{OBS}}}) \sim
\int  p({\Lambda}_{\textsc{\tiny{OBS}}}|a,b,{\Lambda}_{\textsc{\tiny{FLOW}}}) p({\Lambda}_{\textsc{\tiny{FLOW}}})p(b)
\mathrm{d}b \mathrm{d}{\Lambda}_{\textsc{\tiny{FLOW}}}  \ .\label{posteriorTrend}
\end{equation}
Because of the abrupt change in the Earth's oblateness during the $1990$'s, we restrict the 
analysis to pre $1990.0$ epochs.
Observed LOD variations 
${\Lambda}_{\textsc{\tiny{OBS}}}$ are taken from \citet{Gross2001} who also provides uncertainty  estimates
$\Sigma_{{\Lambda}_{\textsc{\tiny{OBS}}}}$. The likelihood
distribution is approximated by a Gaussian distribution such as 
\begin{equation}
({\Lambda}_{\textsc{\tiny{OBS}}}|a,b,{\Lambda}_{\textsc{\tiny{FLOW}}}) 
= \mathcal{N}\left({\Lambda}_{\textsc{\tiny{OBS}}} - a\textbf{t} - {b} - \bar{{\Lambda}}_{\textsc{\tiny{FLOW}}}, \Sigma_{{\Lambda}_{\textsc{\tiny{OBS}}}}\right).
\end{equation}
The prior distribution of ${\Lambda}_{\textsc{\tiny{FLOW}}}$ is also assumed to be Gaussian with a mean and a covariance deriving from the ensemble
of ${\Lambda}_{\textsc{\tiny{FLOW}}}$ time series calculated with equation (\ref{Jault}). The 
distribution $p(a|{\Lambda}_{\textsc{\tiny{OBS}}})$ is thus also a Gaussian distribution. 
Our computation suggest a mean of $\bar{a} = 2.2\,$ms/cy and a large standard deviation of 
$\sigma_a=2.2\,$ms/cy which embraces the values discussed above.

\begin{figure}[h]
\begin{center}
      \includegraphics[width=0.8\linewidth]{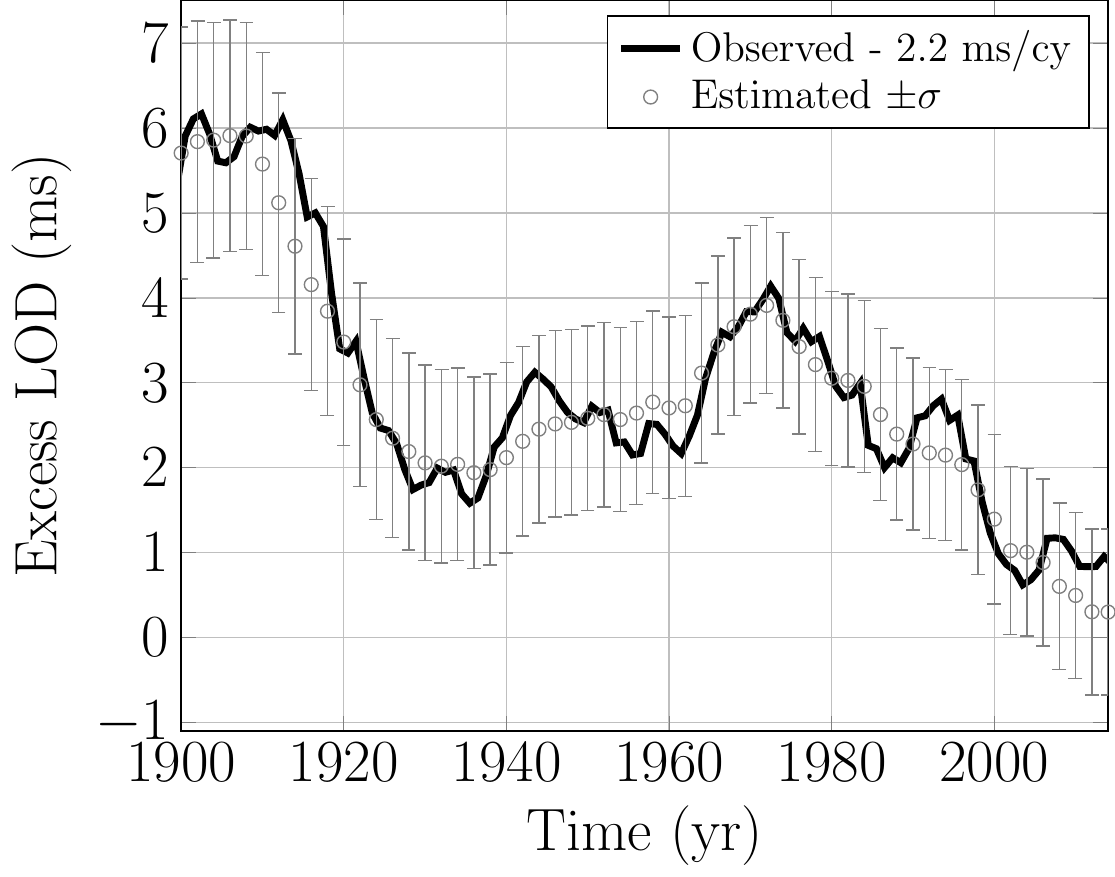}
\caption{Observed variations in length of day (black curve) taken from \citet{Gross2001} and extended after $1997.0$ with a time 
series provided by the Earth orientation center. A trend of $2.2$ms.cy${}^{-1}$ estimated in section 3.4 has been removed
from the observed time series. The variations in length of day induced by the outer core flow and associated standard deviation are shown with
gray circles and error bars.}\label{LOD}
\end{center}
\end{figure}

\noindent The mean estimated trend $\bar{a}$ lies well between the $1.8$ms/cy and $2.4$ ms/cy interval prescribed by tidal forces and by
the measurements of the Earth's oblateness. Furthermore, once the optimal trend is removed from ${\Lambda}_{\textsc{\tiny{OBS}}}$, 
comparisons with the core flow contribution exhibit a good agreement on decadal time scales as illustrated by figure \ref{LOD}.
Although the large uncertainty levels associated with $\bar{a}$ and $\bar{{\Lambda}}_{\textsc{\tiny{FLOW}}}$, 
forbids any precise conclusions on the impact of the last century sea level rise
and ice melting on the variations in length of day, our results suggest that the scenario proposed by \citet{Mitrovica2015} is not the most likely
(otherwise the optimal trend would be close to $1.4$ms/cy).
Instead, estimates of $\dot{\Lambda}_{\textsc{\tiny{GIA}}}$ given by \citet{Peltier2015} are probably more appropriate. This means
that discrepancies between predicted and observed trend in ${\Lambda}$ over the last centuries  could be compensated by a recent
increase in core angular momentum, and that the outer core flow would not slow down the rotation of the Earth over very long periods of time.



\subsection{Predictions}\label{prediction}

\noindent The ability of a model to successfully predict the system evolution 
not only suggests that the model correctly captures the dynamics on the
considered time scales but also points towards useful applications.
We test our model with so-called hindcast simulations, which means that the analysis steps in the assimilation are only 
carried out until a time $T_0$,  
and then integrated as a free model run up to a time $T_F$. The free run corresponds to the forecast, 
or prediction, which can be compared with the data.

\begin{figure}[h]
\begin{center}
      \includegraphics[width=0.99\linewidth]{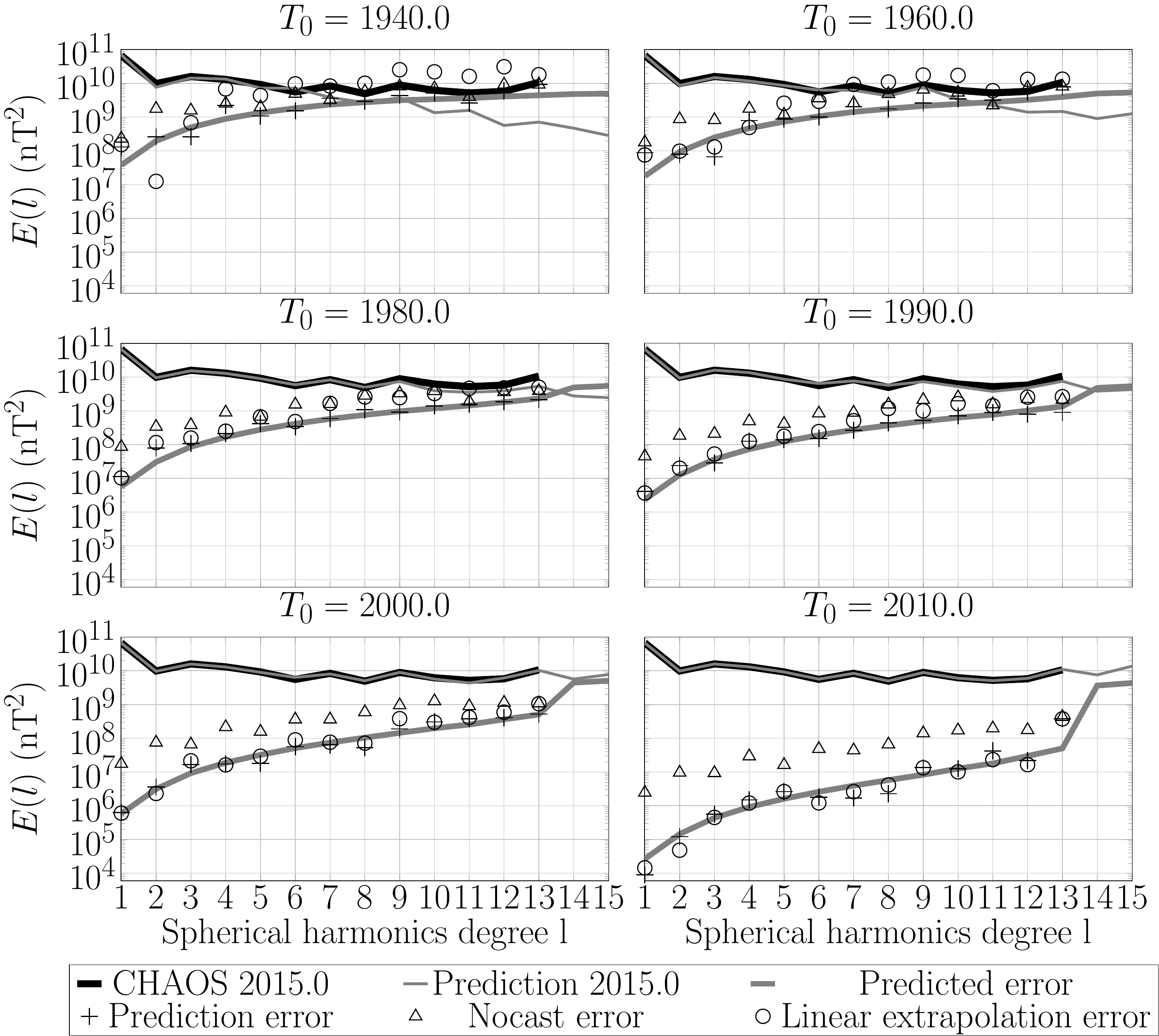}
\caption{Results of the hindcast tests from $T_0$ to $T_F=2015.0$, expressed in terms of energy spectra at the core mantle boundary.
Spectra of the observed magnetic field in $2015.0$ (black lines), its mean prediction (thin gray lines), the prediction error (crosses) and 
the predicted error (thick gray lines).  Triangles and circles respectively correspond to the energy spectra of the linear extrapolation error, and the no cast error.
}\label{predictionSpectra}
\end{center}
\end{figure}

\noindent Here we use six different $T_0$ values, $1940.0$, $1960.0$, $1980.0$, $1990.0$, $2000.0$, and $2010.0$, 
and compare predictions for $T_F=2015$ with the respective epoch in the CHAOS-6 magnetic field model 
by \citet{Finlay2016}. This means that we attempt predictions over periods ranging from $5$ to $55$ years. 
Figure \ref{predictionSpectra} presents the results 
in terms of energy spectra at the Earth's core mantle boundary. 
Thick black lines show the CHAOS-6 reference field $b^c$ while thin gray lines show the 
predicted ensemble mean $\bar{b}^f$. 
Crosses illustrate the prediction error $b^c-\bar{b}^f$ which can be compared with thick gray line that 
depicts the predicted error, which is simply the standard deviation of the $40000$ ensemble members. As expected,
the errors increase with the prediction period (decreasing $T_0$). 
The predicted error provides a good estimate for the prediction error with the exception of 
degree $l=13$ in the $5$ year period. This likely owed to the fact that the CHAOS-6 core field 
contains the crustal field that starts to contribute more significantly at this 
scale. The larger prediction error outweighs this deficiency for longer prediction periods.

\noindent The comparison with two other more trivial prediction methods further allows to judge the 
advantage of our more sophisticated approach. 'No cast' refers to the assumption that the field remains
identical to the field at $T_0$. 'Linear extrapolation' uses the secular variation at $T_0$ to linearly 
extrapolate the field from $T_0$ to $2015$: $b_l(2015) = b^o(T_0) + (2015-T_0)\gamma^o$. The prediction 
errors of these two trivial methods are shown as triangles and circles in figure \ref{predictionSpectra}. 
Linear extrapolation and assimilation prediction errors remain similar for the two shortest 
prediction periods. However, for predictions beyond the $10$ yr horizon, our assimilation formalism starts to 
pay off. For the longest prediction period of $55$ years, the errors in the two trivial methods already exceed 
the spectral energy at degree $6$ while the assimilation predictions remains 
appropriate until degree $9$.

\noindent Working with an ensemble also allows the evaluation of statistical properties of quantities like 
inclination or declination that depend nonlinearly on the state variables. 
Figure 8 compares the
inclination (left) and declination (right) for $1990$ (black lines) and $2015$ (red lines) 
with the ensemble mean predictions using $T_0=1990$ (yellow lines). The prediction errors are 
quantified by the absolute local difference in degrees and are shown as color maps in the top
two panels. Color maps in the bottom two panels show the $90\%$ confidence interval 
of the ensemble prediction.

\begin{figure*}[h]
\begin{center}
   \includegraphics[width=0.8\linewidth]{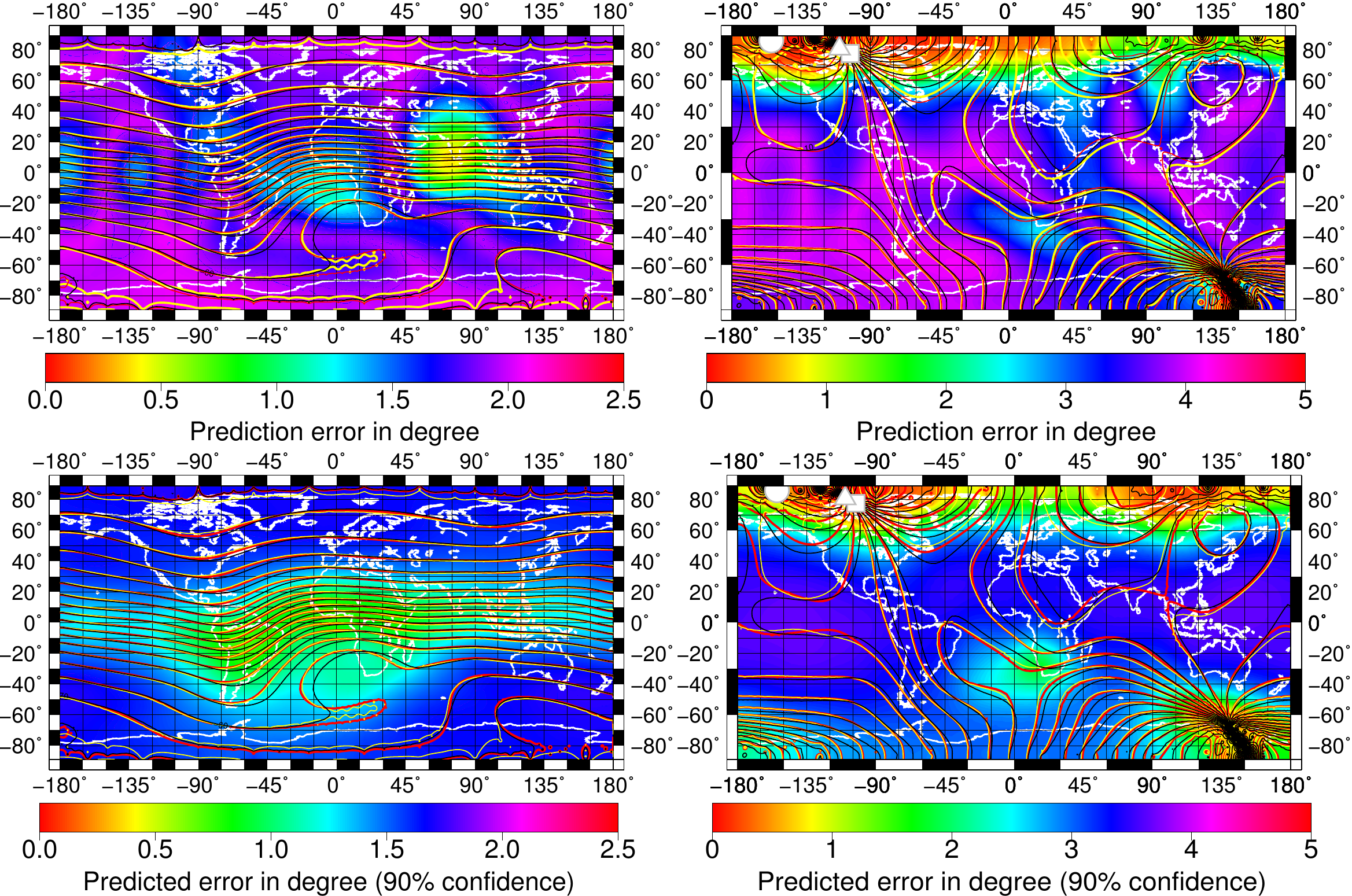}
\caption{$1990.0-2015.0$ hindcast. Isocontours of the observed inclination (left) and  declination (right), in $1990.0$ (black) and $2015.0$ (yellow), and their predictions in 
$2015.0$ (red). Color maps correspond to the absolute value of prediction error (top) and to the $90\%$ confidence interval on the prediction (bottom).
The white symbols on the right are associated with the North magnetic dip pole, observed in $1990.0$ (square) and in $2015.0$ (circle) 
and predicted in $2015.0$ (triangle).}
\label{declinationInclination}
\end{center}
\end{figure*}

\noindent As expected, inclination errors are large for the small or vanishing values around the equator. 
This is illustrated in the top panel of the right column and also well captured by the 
larger variance in the ensemble used to predict the error shown in the lower right panel. 
We calculated the area where the 
prediction error remains within the $90$\% confidence interval defined by the 
prediction ensemble, which amounts to $89.9$\% of the total surface. 
This indicates that the inclination uncertainties provide a good estimate of the prediction error.

\noindent Since the declination becomes undefined at the North and South dip poles, the surrounding regions 
show larger prediction errors and ensemble variances, as is demonstrated in the left column of 
figure 8. 
The area where prediction errors remain within the $90\%$ confidence interval 
amounts to only $81.1\%$ of the surface, which suggests a somewhat inferior error prediction 
likely related to the dip poles. 
The error prediction seems more reliable in the southern hemisphere alone where 
the relative area increases to $92.2\%$.

\noindent Although not clearly visible in figure 8 because the color has been saturated at $5^\circ$, 
the prediction error for the location of the 
North Magnetic Dip Pole (NMDP) is also particularly large. Its real and predicted position for 2015 have 
been marked by gray circles and triangles, respectively. According to \citet{Chulliat2010}, the rapid acceleration of the 
NMDP drift during the 1990's can be explained by the expulsion of magnetic flux below the New Siberian Islands. 
This may be the reason for the larger prediction error, since such expulsions are not modeled in the frozen flux 
approximation used here. Note that the NMDP drift is much better captured when the hindcast test starts at $T_0\geq 2000.0$.

\noindent Finally, we note that our model accurately predicts the evolution of the inclination and declination 
associated with South Atlantic anomaly despite the significant changes between $1990.0$ and $2015.0$.
This highlight the potential usefulness of the method for forecasting core's magnetic field features.

\subsection{Predictability}\label{predictability}

In order to quantify the different sources of forecast errors we analyze the 
secular variation which is responsible for advancing the field from 
$T_0$ to $T_F$: $b(T_F) = b(T_0) + \int_{T_0}^{T_F}\gamma(s) ds$. 
Under the frozen flux approximation, $\gamma$ depends on the flow and the magnetic field, 
through the relation $\gamma = -\nabla_H(ub)$. 
When separating $b$ into the observable part $b^{<}$ and the non observable small 
scale part $b^{>}$ we can distinguish three error sources: 
\begin{equation}
 \gamma^\prime = -\nabla_H(u^\prime b) -\nabla_H(u b^{<\prime}) -\nabla_H(u b^{>\prime}).
\end{equation}
Here the primed quantities on the right hand side indicate deviations from 
the ensemble expectation value, for example $u^\prime = u-\bar{u}$.

\begin{figure}[h]
\begin{center}
      \includegraphics[width=0.99\linewidth]{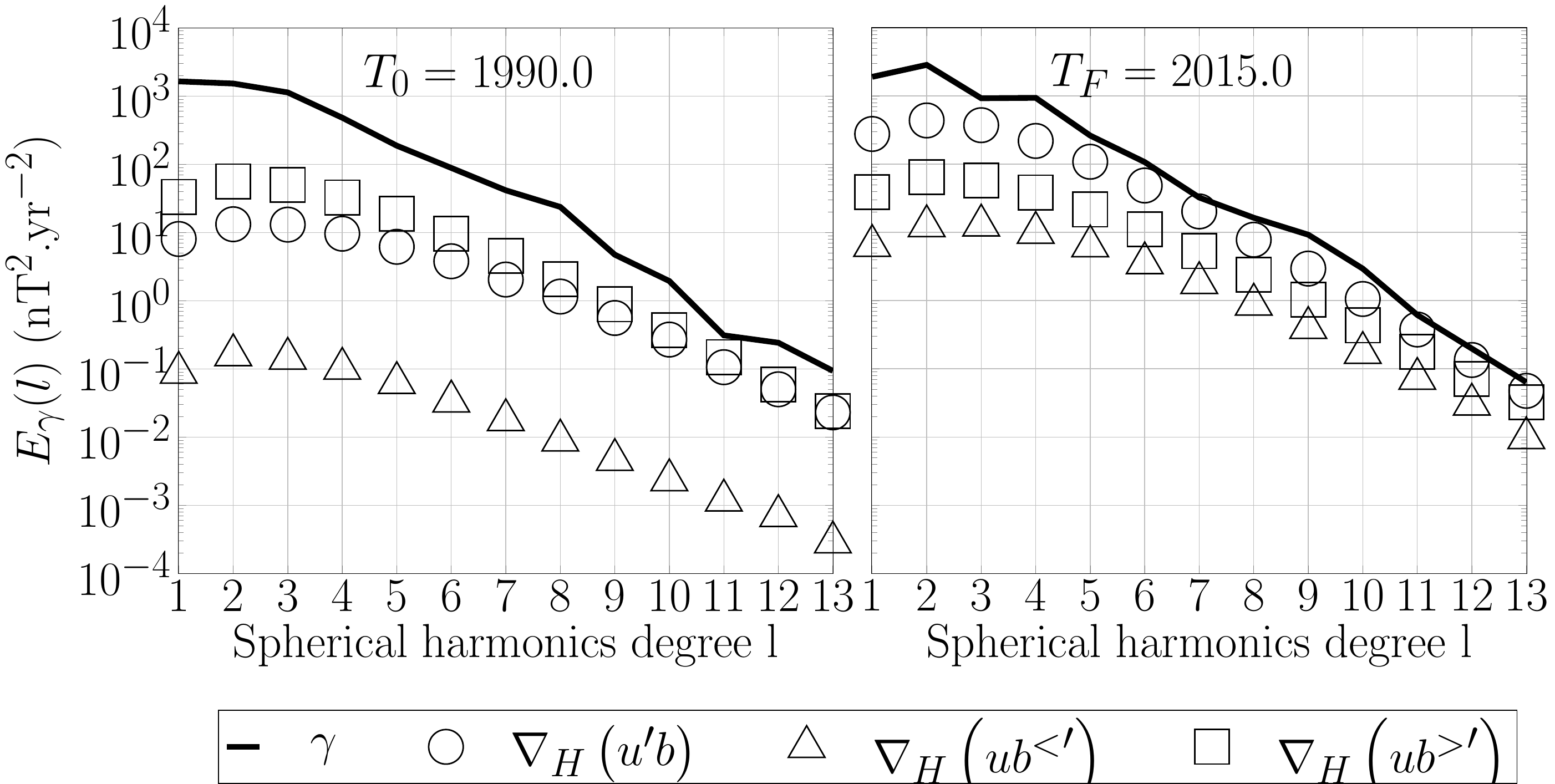}
\caption{Sources of secular variation (SV) error in magnetic field predictions, at the initial time $T_0=1990.0$ (left) and at the forecast time $T_F=2015.0$(right).
Energy spectra at the Earth's surface of the secular variation  (black lines) and of different types of error (symbols) generated
when estimating the SV. Triangles and  circles 
correspond to the errors induced by respectively the large scale and small scale variable parts of the magnetic field.
Squares are associated with the uncertainties arising from the variable part of the velocity field. 
}\label{SVuncertainties}
\end{center}
\end{figure}

Figure 9 compares the spectra of the different error contributions to the 
secular variation spectrum at the Earth's surface for $T_0=1990$ (left) and 
$T_F=2015$ (right). 
The last analysis step performed at $T_0$ directly constraints $b^{<}$ and 
leads to a small related variance and thus a small error contribution that lies
about two orders of magnitude below the $u$ and $b^{>}$ related error. 
At the end of the forecast at $2015$, however, the 
$b^{<}$ related error has grown significantly but remains the smallest contribution.
Since $b^{>}$ is mostly constrained by the apriorily assumed statistics at all times, 
the related error changes only mildly. While the $u$ related error 
is smaller than the $b^{>}$ error at $1990$, it becomes the dominant 
contribution at $2015$ due to the increase in flow dispersion.

\noindent Thus neither the observational error in the large scale magnetic field nor the 
lack of knowledge on the small scale contributions is the limiting factor for the 
predictions but rather randomization of the different velocity field scales over time.
\noindent We also tested the predictability range for the magnetic field. Starting a forecast 
in $2014$, we let the system evolve until the scale per scale variance of the magnetic field exceeds the mean predicted energy. 
Like in the forecast from 1990 to 2015 illustrated in the middle left panel of figure  \ref{predictionSpectra}, after $75$ yrs only, the variance 
in contributions beyond degree $l=7$ exceed the mean energy. The 
predictability limit further decreases to $l=5$, $l=3$, and $l=2$ after $160$yrs, $400$yrs and $640$ yrs repsectively. 
After $1950$ yrs, even the dipole energy is exceeded by the respective variance level.

\section{Conclusion}\label{conclusion}

\noindent We have employed a sequential data assimilation framework to model the dynamics of the geomagnetic field  and the flow at the top of 
Earth's core in the 20th century using the COV-OBS.x1 model of \citet{Gillet2015} as observations. 
The method extends the approach in \citet{Baerenzung2016} to the time domain, as a sequential propagation in time of flow and field 
inversions under weak prior constraints.
The prior is a dynamical model that combines the induction equation in the frozen flux approximation with a simple 
AR1 process describing the flow evolution.
The latter comprises a memory term and stochastic forcing which are both constrained by the secular variation 
observations following the ideas presented in \citet{Baerenzung2016}.

\noindent We use an ensemble approach, the EnKF (\citet{Evensen2003}), where the dynamical model uncertainties are characterized by statistically sound covariances. 
Using the AR1 process falls short of integrating a proper Navier-Stokes equation but allows us to forward a large ensemble 
of 40,000 members in time in order to characterize the errors and prior covariances. 

\noindent The optimal parameters characterizing the flow prior spatial and temporal properties points to particularly long time scales 
of several centuries to millennia for the toroidal field at SH degrees $l=1$ and $2$. 
This flow contribution can be attributed to a large scale slowly evolving gyre which has also been identified in core flow inversions 
(\citet{Pais2000,Gillet2015}) and numerical simulations (\citet{Aubert2013b,Schaeffer2017}).
The most prominent feature of the gyre is the well documented pronounced westward drift at low and mid-latitudes of the Atlantic hemisphere. 
Smaller scale flows have characteristic time scales in the decadal range that are consistent with previous estimates (\citet{Christensen2004,Hulot2010}). 
Typical related features are local modifications of the gyre in the southern hemisphere or the acceleration of the westward 
flow at and around the tangent cylinder underneath Alaska and the eastern part of Siberia, which has already been reported by \citet{Livermore2017}.
This points to an important contribution of ageostrophic motions to the dominantly geostrophic overall core flow.

\noindent Predictions of decadal length of day variations (LOD) from changes in angular momentum of our core flow model yield remarkable resemblance 
to corresponding independent observations. Furthermore, the recent increase in core angular momentum, that we attribute to an acceleration
of the geostrophic contribution of the gyre, enables to compensate the difference between
the recently observed trend in LOD changes and the expected one highlighted by \citet{Munk2002}.

\noindent We further tested the capability of our model to forecast the evolution of magnetic and flow fields through hindcast experiments. 
Comparisons of the magnetic field evolution with linear extrapolations and no casts (in which the field is assumed static) show that our more 
sophisticated model significantly improves predictions beyond 10 or 15 years. 
Moreover, we inspected the reliability of the forecast errors predicted by the ensemble dispersion, which resulted in good agreement with the hindcast errors.
Such match reveals that the characteristic times estimated for the auto regressive process of the flow correspond to a realistic randomization rate of the fields.

\noindent However, it is the dispersion of the velocity field itself which seems to dominate the uncertainties in secular variation estimations,
what limits therefore the predictability of the geomagnetic field. 
Within this limitation, the scale dependent predictability corresponds to 1950, 640, 400, 160 and 75 years for degrees $l=1$, 2, 3, 5 and 7, respectively. 
A more realistic dynamical model such as a geodynamo simulation would possibly extend the predictability horizon. 
Nevertheless, the enormous 
numerical power required to perform dynamo simulations at more extreme parameters 
would preclude the type of ensemble Kalman filter approach followed here.
Even with compromises in dynamo model parameters and the ensemble size, the 
computational costs would still increase by orders of magnitude. 
Moreover, since dynamo simulations are strongly nonlinear, the system bears an intrinsic sensitivity to initial perturbations. 
This amounts to an important e-folding time, which is estimated to be about 30 years with a temporal rescaling within the secular variation 
time scale (\citet{Hulot2010,Lhuillier2011}).
Since this characteristic time is not so different from the one associated with the small length scales of our flow model, 
we expect that the rate at which information is lost in the system will be somewhat equivalent in both modeling strategies.

\appendix
\section{Gibbs sampling}\label{appA}

\noindent The Gibbs sampling algorithm permits to randomly draw an ensemble characterizing statistically a given joint distribution, by recursively 
sampling conditional probability distributions deriving from it. In our case, the joint distribution of interest is $p(\bu,\bb|\gamma_o)$.
Therefore, at a step $n$, the algorithm samples alternatively the two following distributions:
\begin{eqnarray}
 p(\bu^n|\bb^{n\mbox{-}1},{\bf \gamma}_{o},\hat{\MM}) & = &\mathcal{N}\left( \bar{\bu}^n,{\bf \Sigma_{u^{\textnormal n}}} \right) \label{gibbsConditonnedU} \\
 p(\bb^n|\bu^{n},{\bf \gamma}_{o},\hat{\MM}) & = &\mathcal{N}\left( \bar{\bb}^n,{\bf \Sigma_{b^n}} \right) \label{gibbsConditonnedB}
\end{eqnarray}
with:
\begin{eqnarray}
 \bar{\bu}^n & = & \bar{\bu} + {\bf \Sigma_{u_{|\hat{\MM}}}}  {\bf A_{b^{n\mbox{-}1} }^T} {\bf R_{u^n}^{-1}} 
 \left( \bar{{\bf \gamma}}^o + {\bf A_{b^{n\mbox{-}1}}}\bar{\bu} \right) \\
{\bf \Sigma_{u^{\textnormal n}}} & = &  {\bf \Sigma_{u_{|\hat{\MM}}}}  -  {\bf \Sigma_{u_{|\hat{\MM}}}}  {\bf A_{b^{n\mbox{-}1} }^T} {\bf R_{u^n}^{-1}} 
{\bf A_{b^{n\mbox{-}1} }} {\bf \Sigma_{u_{|\hat{\MM}}}} \\
 \bar{\bb}^n & = & 
 \bar{\bb}  + {\bf\Sigma_{b}}{\bf A_{u^{n}}^T} {\bf R_{b^n}^{-1}} 
 \left( \bar{{\bf \gamma}}^o + {\bf A_{u^{n}}}\bar{\bb}  \right) \\
{\bf \Sigma_{b^n} }& = & {\bf \Sigma_{b} } -  {\bf\Sigma_{b}}{\bf A_{u^{n}}^T} {\bf R_{b^n}^{-1}} {\bf A_{u^{n}}} {\bf\Sigma_{b}} \label{gibbsVarianceB}\ .
\end{eqnarray}
and where:
\begin{eqnarray}
{\bf R_{u^n}}& = &\left(  {\bf A_{b^{n\mbox{-}1} }} {\bf \Sigma_{u_{|\hat{\MM}}}}  
 {\bf A_{b^{n\mbox{-}1} }}^T + {\bf \Sigma_{\gamma}^o} \right) \\
{\bf R_{b^n}}& = &\left(  {\bf A_{u^{n} }} {\bf \Sigma_{b}}  
 {\bf A_{u^{n }}^T} + {\bf \Sigma_{\gamma}^o} \right) \ .
\end{eqnarray}

\noindent In this study, three epochs were considered simultaneously, $1900.0$, $1950.0$ and $2000.0$.
Since over long periods of time the real temporal correlations of the slow varying components of the flow probably differ from 
the one induced by the auto regressive process, the characteristic times associated with the degree $l=1$ and $l=2$ of the toroidal field and 
$l=1$ of the poloidal field were reestimated with a $50$ year time step.

\begin{acknowledgments}
This work has been supported by the German Research Foundation (DFG) within the Priority Program
SPP1788 ``Dynamic Earth''. The data for this paper are available by contacting the corresponding author at baerenzung@gmx.de.
\end{acknowledgments}

%
%

\end{article}

\end{document}